\documentclass[preprint,aps,eqsecnum,nofootinbib]{revtex4}
\usepackage{amsmath, amssymb}
\usepackage{fancybox}
\usepackage{graphicx}
\usepackage{dcolumn}
\usepackage{bm}
\usepackage[dvips]{color}
\usepackage{wrapfig}
\newcommand{\half}{\frac{1}{2}}

\newcommand{\der}{\partial}

\newcommand{\Tr}{\mbox{\rm Tr}}

\newcommand{\bfk}{\bm{k}}

\newcommand{\bfp}{\bm{p}}

\newcommand{\bfu}{\bm{u}}

\newcommand{\bfx}{\bm{x}}

\newcommand{\bfP}{\bm{P}}

\begin{document}

\preprint{KYUSHU-HET-95}

\title{Anomalous dimensions determine the power counting \\
--- Wilsonian RG analysis of nuclear EFT ---}

\author{Koji Harada}
\email{koji1scp@mbox.nc.kyushu-u.ac.jp}
\author{Hirofumi Kubo}
\email{kubo@higgs.phys.kyushu-u.ac.jp}
\affiliation{Department of Physics, Kyushu University\\ 
Fukuoka 812-8581 Japan}

\date{\today}

\begin{abstract}
 The Legendre flow equation, a version of exact Wilsonian
 renormalization group (WRG) equation, is employed to consider the power
 counting issues in Nuclear Effective Field Theory. A WRG approach is an
 ideal framework because it is nonperturbative and does not require any
 prescribed power counting rule. The power counting is determined
 systematically from the scaling dimensions of the operators at the
 nontrivial fixed point. The phase structure is emphasized and the
 inverse of the scattering length, which is identified as a relevant
 coupling, is shown to play a role of the order parameter. The relations
 to the work done by Birse, McGovern, and Richardson and to the
 Kaplan-Savage-Wise scheme are explained.
\end{abstract}

\pacs{Valid PACS appear here}
\maketitle

\section{Introduction}


Conventional nuclear theory is based on force potentials.  There have
been much progress and we now have very accurate ones such as
Nijmegen\cite{Stoks:1994wp}, Argonne $V_{18}$\cite{Wiringa:1994wb}, and
CD-Bonn\cite{Machleidt:1995km,Machleidt:2000ge} potentials which fit
nicely to about 3,000 nucleon-nucleon scattering data with energies up
to 350 MeV. The potential models have matured to the demand of precise
numerical calculations.


Potential models however have serious drawbacks. Even though they are
very precise, after all, they are just (semi-phenomenological)
models. It is not obvious how these models are related to QCD, the
fundamental theory of strong interactions, nor how to improve them in a
systematic manner.  Proper treatment of inelastic processes seems
formidable. In addition, the potential approach is known to suffer from
the so-called \textit{off-shell ambiguities}, which may cause serious
problems for a correct understanding of multi-nucleon systems.


Nuclear effective field theory (NEFT)\cite{Weinberg:1990rz,
Weinberg:1991um} is expected to be a promising alternative. It is based
on very general principles of quantum field theory\cite{Weinberg:1978kz,
Lepage:1989hf, Polchinski:1992ed, Georgi:1994qn, Kaplan:1995uv,
Manohar:1996cq}, and linked to QCD through chiral symmetry, while,
instead of quarks and gluons, only the relevant degrees of freedom
(nucleons, and optionally pions, deltas, etc.)  at low energies are
considered. Chiral perturbation theory($\chi$PT)\cite{Gasser:1983yg,
Gasser:1984gg}, a prominent example of EFT in which only meson degrees
of freedom are treated explicitly, has been applied very
successfully. In short, NEFT aims at a similar success.


Since a Lagrangian of EFT contains an infinite number of operators
allowed by the symmetry, one needs a power counting rule to give
orderings to an infinite number of Feynman diagrams generated by the
Lagrangian.  Once it is specified, physical quantities can be calculated
in a systematic expansion in powers of $Q/\Lambda_0$, where $\Lambda_0$
is the cutoff, above which the EFT is not valid, and $Q\ (< \Lambda_0)$
is the typical momentum of the process one is interested in.  In the
case of perturbative systems such as meson-meson scattering, the power
counting which is determined on the basis of naive dimensional
analysis(NDA)\cite{Manohar:1983md} is known to give correct
results. NEFT is, on the other hand, inherently nonperturbative. (There
are bound states, the nuclei!) In addition, it is a fine-tuned system;
the inverses of the nucleon-nucleon scattering lengths $1/a^{({}^1S_0)}
\sim-8$ MeV, and $1/a^{({}^3S_1)} \sim 40$ MeV are much smaller than
$\Lambda_0\sim m_\pi \sim 140$ MeV for the pionless NEFT. (If we
consider the pionful NEFT, $\Lambda_0$ should be larger, presumably $
\Lambda_0 \sim 500$ MeV, and the system should be considered more
fine-tuned.)  Moreover deuteron is a shallow bound state with binding
energy of 2.2 MeV. We need a power counting scheme which takes into
account the ``unnaturalness.''


In this paper, we consider a systematic way of determining the power
counting for such nonperturbative and fine-tuned systems on the basis of
Wilsonian RG analysis. As we will explain below, our approach is very
general and not restricted only to NEFT.


At present, there are basically two power counting schemes proposed for
NEFT. In Weinberg's scheme\cite{Weinberg:1990rz, Weinberg:1991um}, one
identifies the sum of irreducible diagrams in the expansion based on NDA
as the ``effective potential,'' and solves the Lippmann-Schwinger (L-S)
equation with it. The ``effective potential'' is iterated to all order,
i.e., all operators are treated nonperturbatively. Although conceptually
attractive and simple in numerical implementation, Weinberg's power
counting has been shown to be inconsistent: the divergences that arise
in the leading order calculation cannot be absorbed by the leading order
operators in certain partial waves\cite{Kaplan:1996xu, Nogga:2005hy}.
The other power counting scheme proposed by Kaplan, Savage and Wise
(KSW)\cite{Kaplan:1998tg, Kaplan:1998we} begins with a renormalization
group (RG) behavior at a nontrivial fixed point, which corresponds to a
system with infinite scattering length.  In contrast to Weinberg's
scheme, only the nonderivative contact operator is treated
nonperturbatively while others are treated perturbatively.  Although the
KSW power counting is a consistent scheme in the sense that results are
independent of subtraction point at every order of the expansion,
serious problems have been reported.  Fleming, Mehen, and Stewart
(FMS)\cite{Fleming:1999ee} calculated the nucleon-nucleon scattering
amplitudes up to next-to-next-to leading order in a NEFT with pions
included (``pionful'' NEFT), implementing the KSW power counting, and
found that the scattering amplitude in the ${}^3S_1$--${}^3D_1$ channel
shows no convergence at all. A possible explanation for the success and
the failure has been given in Ref.\cite{Beane:2001bc,Birse:2005um}.


The present status may be summarized as that NEFT works ``pretty well if
one follows a patchwork of power counting rules\cite{Kaplan:2005es}.''
Actually, in spite of the problems mentioned above, many of higher order
calculations in both power counting schemes have been found to be
successful at least numerically. It might be however too early to say
that the problems are settled. One should keep in mind that, besides the
numerical agreement with data, a thorough understanding of power
counting in a systematic and consistent way is extremely important.
It is this issue that we address in this paper.


We believe that Wilsonian Renormalization Group(WRG)\cite{Wilson:1973jj,
Wilson:1974mb} approach can provide the key.  This approach allows us to
analyze the behavior of the operators (or the corresponding coupling
constants) by assuming only the symmetry and the relevant degrees of
freedom. Most importantly to our present purpose, WRG analysis is
nonperturbative and does not require any prescribed power counting
scheme.


Our basic assertion is that it is anomalous dimensions that determine
the power counting.  We are going to show in this paper that it is very
natural to determine the power counting of an operator from the
corresponding scaling dimension (the sum of the canonical and the
anomalous dimensions). Our reasoning is based on the very general idea
that the power counting should be determined by dimensional
analysis. Furthermore, our WRG approach is applicable both to trivial
and nontrivial fixed points; near the trivial fixed point, it coincides
with NDA, while it explains why the KSW power counting is successful to
some extent near the nontrivial fixed point. The real two-nucleon system
happens to be close to the nontrivial fixed point, but the WRG approach
has a wider scope.


Wilsonian approach has been applied by Birse, McGovern, and
Richardson\cite{Birse:1998dk} to nonrelativistic two-body scattering to
obtain a power counting rule.  They considered a Wilsonian RG equation
for an effective potential, and identified two fixed points; a trivial
one and a nontrivial one.  Then they determined how to organize the
terms in the effective potential by considering the linearized RG
equation around the fixed point.  Although the potential they considered
is general, the physical meaning of energy-dependence of it \textit{in
the Schr\"odinger equation} is obscure.  Furthermore it does not seem to
be very clear how the symmetry of the theory is implemented in the
potentials.  Since this potential approach is similar to traditional
potential model approach, it might share some of their problems.  It is
desirable to understand the power counting in a completely field
theoretical framework to benefit from all the good features of EFT.  The
relation between their work and the present work will be explained in
Sec.~\ref{sec:comments} in detail.


A WRG analysis in field theory requires all possible operators
consistent with the assumed symmetry. In the previous
paper\cite{Harada:2005tw}, we emphasized the role played by ``redundant
operators,'' which may be eliminated by field redefinitions. We then
explicitly showed that the nonperturbative off-shell amplitude can be
renormalized only by properly treating redundant operators in the WRG
analysis. However, the method of the calculation is rather
straightforward and seems applicable only for simple cases. A more
powerful method is needed for general cases.


In this paper we study the RG flow for the pionless NEFT employing
Legendre flow equation, which is one of the implementations of Wilsonian
RG. We obtain the RG flows for ${}^1S_0$ and ${}^3S_1$--${}^3D_1$
channels to order $\mathcal{O}(p^2)$.  We identify two phases in our RG
flow for both channels; the strong coupling phase and the weak coupling
phase.  We also show that the nontrivial fixed point is on the phase
boundary of our RG flow.  The scaling dimension of the operators around
the fixed points are calculated. We find that the anomalous dimension of
the operators around the nontrivial fixed point is large. We argue that
the scattering length is identified with the order parameter which
characterizes these phases.


The pionless NEFT is an interesting theory as an application of
nonperturbative RG equation, which is, in most cases, very difficult to
treat. Thanks to the nonrelativistic feature of the theory, it allows us
to solve (approximately) with a simple truncation of the space of
operators, though the justification of the truncation only comes from
actually enlarging the space. This theory also provides a simple example
of nontrivial fixed point and the existence of a bound state. We hope
that analyzing this theory gives some insight into the use of
nonperturbative RG to get the information about bound states.


This paper is organized as follows: in Sec.~\ref{sec:powercounting}, we
explain why the scaling dimensions are important in determining power
counting.  The Legendre flow equation for nonrelativistic systems is
introduced in Sec~\ref{sec:Legendre}.  We present the detailed analysis
of pionless NEFT and its physical implications in
Sec.~\ref{sec:pionless}. The relation to the work done by Birse
\textit{et al.} is clarified and comments on the other schemes,
especially on the KSW scheme, are made in
Sec.~\ref{sec:comments}. Finally in Sec.~\ref{sec:summary}, we summarize
our study, and discuss future prospects. Appendix~\ref{sec:ir} provides
some technical information about the cutoff function used in the
Legendre flow equation. In Appendix~\ref{sec:amplitudes} we derive the
RG equations from the two-nucleon scattering amplitudes.

\section{Power counting and Renormalization Group}
\label{sec:powercounting}

In this section, we consider what power counting is and its relation to
renormalization group in general terms.


The most basic idea behind power counting is the order of magnitude
estimate based on dimensional analysis. The order of magnitude of a
physical quantity may be estimated by an appropriate combination of
dimensionful parameters of typical scales. The period $T$ of a simple
pendulum with length $L$ and mass $M$ in a uniform gravitational field
with the gravitational acceleration $g$ is estimated as $T \sim
\sqrt{L/g}$. The dimensionless coefficient is expected to be of order
one, the idea of ``naturalness.'' (In the pendulum example, it is a bit
large, $2\pi$.)


There is only one important scale in a simplest (relativistic) field
theory; the physical cutoff $\Lambda_0$. The mass $m$ is related to
$\Lambda_0$ by a very large dimensionless constant $\xi$ as
$m=\Lambda_0/\xi$. (This is a necessary fine-tuning to have a sensible
quantum field theory.)  Any quantities may be measured in units of
$\Lambda_0$. The interaction Lagrangian in spacetime dimension $D$ may
be written as an infinite sum of the interactions of the form
\begin{equation}
 \mathcal{L}_{int}=-\sum_i \frac{g_{0i}}{\Lambda_0^{d_i-D}} \mathcal{O}_i(x),
\end{equation}
where $d_i$ is the (canonical) dimension of the operator
$\mathcal{O}_i(x)$ and $g_{0i}$ is the (bare) dimensionless coupling
constant. (It is sometimes useful to say that the (dimensionful)
coupling $\tilde{g}_{0i}\equiv g_{0i}/\Lambda_0^{d_i-D}$ has dimension
$D-d_i$.)  Classically (and near the trivial fixed point, see below)
this dependence on $\Lambda_0$ determines how large the contribution of
$\mathcal{O}_i$ is. One may expand the contribution in powers of
$(Q/\Lambda_0)$, where $Q$ is the typical energy/momentum scale of
interest, according to the canonical dimensions of the operators. This
is a common situation, and, with a twist of chiral symmetry, true for
$\chi$PT\footnote{In $\chi$PT, the masses of the Nambu-Goldstone bosons
play a special role as symmetry breaking parameters, which are not
related to the cutoff $\Lambda_0$. It is not trivial how to treat
them. The standard way is to treat $m^2\sim Q^2$, which is necessary to
be consistent with the meson pole structure of the
amplitudes\cite{Gasser:1983yg}. In the pionless NEFT, which we consider
in this paper, there is no such a symmetry breaking parameter (we ignore
isospin breaking), and thus no subtlety associated to it.}. The
so-called NDA applies to such cases. In some cases, however, quantum
fluctuations change the situation drastically.


In a WRG analysis, the cutoff is lowered to $\Lambda < \Lambda_0$ by
integrating out the fluctuations with momentum $\Lambda < p <
\Lambda_0$, and thus the interaction Lagrangian may be replaced by
\begin{equation}
 \mathcal{L}_{int}
  =-\sum_i \frac{g_i(\Lambda)}{\Lambda^{d_i-D}} \mathcal{O}_i(x).
\end{equation}
The coupling $g_i(\Lambda)$ now depends on the ``floating'' cutoff
$\Lambda$ so that the physical quantities do not depend on $\Lambda$.
The behavior of $g_i(\Lambda)$ contains information about how the
quantum fluctuation modifies the importance of the operator
$\mathcal{O}_i$. The differential equations,
\begin{equation}
\frac{dg_i}{dt}=\beta_i(g),\qquad
t \equiv \ln\left(\frac{\Lambda_0}{\Lambda}\right),
\end{equation}
are called \textit{RG equation}, and determine the \textit{RG flow} in
the space of all coupling constants.


An important concept in the RG analysis is a fixed point of the RG flow,
which is a solution of $\beta_i(g^{\star})=0$ for all $i$, where the
coupling constants stop running.  Suppose we are interested in the
behavior of a theory close to a fixed point.  The behavior of the RG
flow near a fixed point can be examined by considering the linearized RG
equation obtained by substituting $g_i=g^{\star}_i+\delta g_i$ into the
RG equation keeping only the linear terms in $\delta g_i$,
\begin{equation}
\frac{d}{d t}\delta g_i= A_{ij}(g^{\star})\delta g_j,
\end{equation}
where $A_{ij}(g^{\star})\equiv (\partial \beta_i/\partial g_j)|_{g^{\star}}$.
By diagonalizing $A_{ij}(g^\star)$, we have
\begin{equation}
\frac{d\bfu_i}{d t}=\nu_i \bfu_i, \label{scalingdim}
\end{equation}
where $\nu_i$ is the eigenvalue and $\bfu_i$ is the corresponding
eigenvector, which may be immediately integrated as,
\begin{equation}
 \bfu_i(\Lambda)
  =\bfu_i (\Lambda_0)\left(\frac{\Lambda}{\Lambda_0}\right)^{-\nu_i}.
\end{equation}
The exponent of $\Lambda_0$, $\nu_i$, is called \textit{scaling
dimension} of the coupling.  Depending on the sign of the scaling
dimension, interactions are divided into two groups.  An operator with
its coupling having $\nu_i<0$ is called \textit{irrelevant}.  On the
other hand, that with $\nu_i>0$ is called \textit{relevant}. If
$\nu_i=0$, the operator is called \textit{marginal}. The scaling
dimension is the quantum counterpart of the (canonical) dimension. At
the trivial fixed point, $g_i^\star = 0$ for all $i$, it agrees with the
canonical one, $\nu_i=D-d_i$, i.e.,
\begin{equation}
 \frac{g_i(\Lambda)}{\Lambda^{d_i-D}}\sim \frac{g_{0i}}{\Lambda_0^{d_i-D}},
\end{equation}
while at a nontrivial fixed point it may be quite different. It should
be emphasized that the scaling dimensions are not prescribed but
determined by the theory (and the fixed point) itself. This is the
beauty of the WRG approach.


Let us consider what happens near a nontrivial fixed point more in
detail. Suppose that coupling constant $g_i(\Lambda)$ is written as
\begin{equation}
 g_i(\Lambda)\sim g^\star_i
  +\sum_k c_{ik}\left(\frac{\Lambda}{\Lambda_0}\right)^{-\nu_k},
  \label{g-gstar}
\end{equation}
where $c_{ik}$ is a small constant.  First of all, it is important to
note that the theory is scale invariant at the fixed point. It means
that the theory has no typical scale, though it starts with the physical
cutoff $\Lambda_0$. Actually, the coupling
$g_i(\Lambda)/\Lambda^{d_i-D}\sim g_i^\star/\Lambda^{d_i-D}$ is
independent of $\Lambda_0$. Therefore, \textit{the value of the fixed
point itself does not contribute to the power counting}, because, as we
emphasized above, power counting is based on dimensional
analysis, but a scale invariant theory does not have a scale! It is the
$\Lambda_0$ dependent part (the second term in (\ref{g-gstar})) that
contribute to the power counting.
\begin{equation}
 \frac{g_i(\Lambda)}{\Lambda^{d_i-D}}\sim \frac{g_i^\star}{\Lambda^{d_i-D}}
  +\sum_k c_{ik}
  \frac{\Lambda_0^{\nu_k}}{\Lambda^{d_i-D+\nu_k}}.
  \label{nearnontrivial}
\end{equation}
From the dependence on $\Lambda_0$, one sees that the $k$-th term in the sum
behaves like a coupling with dimension $\nu_k$. 


In the literature, many authors seem to think that the lowest order
contact operator in NEFT should be treated ``nonperturbatively,''
because the value of the corresponding coupling constant is large at the
nontrivial fixed point (the first term of (\ref{nearnontrivial})). As we
will show in Sec.~\ref{sec:pionless}, however, it should be treated
``nonperturbatively'' \textit{because it is relevant} at the nontrivial
fixed point. We believe that this difference is crucial in understanding
the power counting for other operators.


Let us summarize what we explained in this section; power counting is in
essence the order of magnitude estimate based on dimensional
analysis. But the quantum fluctuations change the dimension drastically
in some cases. A WRG analysis takes into account the quantum
fluctuations in a controlled way, and gives rise to a RG flow, which is
characterized by fixed points. Especially near a nontrivial fixed point,
the scaling dimension may be quite different from the canonical one. It
is therefore natural to consider the power counting based on the quantum
scaling dimensions. The important point is that it is not the value of
the nontrivial fixed point, but the scaling dimensions that contribute
to the power counting.


Finally, we would like to make a comment on the use of momentum cutoff
in EFT. Although it might not be widely understood, the momentum cutoff
scheme \textit{does not} break the EFT expansion. (Of course it is much
simpler with dimensional regularization.) The loop contributions which
(partially) cancel the $\Lambda_0$ in the denominator may be absorbed in
the renormalized couplings of lower order. Thus, when expressed in terms
of renormalized couplings, EFT with momentum cutoff has a sensible
expansion as that with dimensional regularization.

\section{Legendre flow equation for non-relativistic theory}
\label{sec:Legendre}


Legendre flow equation\cite{Wetterich:1989xg, Wetterich:1992yh,
Berges:2000ew}, which we use in our analysis, is one of the
implementations of WRG equation.  It is formulated as a RG equation for
the infrared (IR) cutoff effective action called \textit{effective
average action}.  In contrast to the conventional effective action in
which all the fluctuations are taken into account, the effective average
action includes only the quantum fluctuations with momenta larger than
some cutoff $\Lambda$.  An infinitesimal change of an effective average
action with respect to an infinitesimal change in $\Lambda$ is expressed
in a differential equation.  Starting from an arbitrary bare action at
ultraviolet (UV) scale $\Lambda_0$, the effective averaged action
successively includes the lower momentum fluctuations as the cutoff is
lowered, approaches the conventional effective action in the limit of
$\Lambda=0$.


In the case of relativistic theory, it is necessary to define the flow
equation in Euclidean space because a cutoff must be imposed on all four
components of the momenta in order to respect the Lorentz invariance.
In our case of nonrelativistic theory, on the other hand, corresponding
symmetry is Galilean invariance and rotation symmetry.  There is no
obvious way of imposing a cutoff maintaining them. It is a rather
annoying issue in applying the method to nonrelativistic
cases. Fortunately, as we explain in the next section, the correct way
of implementing a cutoff is clear in the two-nucleon system. In this
section, however, we naively consider the flow equation defined in
Minkowski spacetime and impose a cutoff only on three-momenta, though it
breaks Galilean invariance.


We begin by defining IR cutoff generating functional for nonrelativistic
nucleons with source terms, $\eta^{\dagger}$ and $\eta$.  The generating
functional for this system is defined in terms of integration over
Grassmann variables, $N(x)$, by
\begin{eqnarray}
e^{i W_{\Lambda}[{\eta}^{\dagger},\eta ]}= 
\int \mathcal{D}N\mathcal{D}{N}^{\dagger}\ 
e^{i\left(S_{\Lambda_0}
+N^{\dagger}\cdot R^{(1)}_{\Lambda}\cdot N
+{\eta}^{\dagger}\cdot N
-N^{\dagger} \cdot\eta\right)},
\label{generating}\\
N^{\dagger}\cdot R^{(1)}_{\Lambda}\cdot N\equiv \int\frac{d^4p}{(2\pi)^4}
\left(N^{\dagger}(p)\right)_A\left(R^{(1)}_{\Lambda}(p)\right)_{AB}
\left(N(p)\right)_B,
\end{eqnarray}
where $S_{\Lambda_0}$ is an arbitrary bare action composed of local
operators at physical UV cutoff scale $\Lambda_0$.  The indices $A$ and
$B$ denote spin and isospin collectively.  Fourier transform is
defined as follows,
\begin{equation}
N(x)\equiv \int\frac{d^4p}{(2\pi)^4}e^{-ip\cdot x}N(p),\quad
N^{\dagger}(x)\equiv \int\frac{d^4p}{(2\pi)^4}e^{+ip\cdot x}N^{\dagger}(p).
\end{equation}
The function $R^{(1)}_{\Lambda}(p)$ effectively cuts off the IR part of
the fluctuations in the integrand so that $W_{\Lambda}[\eta^\dagger,
\eta]$ contains the effects of only the fluctuations for
momenta larger than $\Lambda$  in the coupling constants.
In principle, $R_{\Lambda}(p)$ can be an arbitrary function
which satisfies the following properties,
\begin{equation}
\left|R^{(1)}_{\Lambda}(\bfp^2)\right| \to
\left\{
 \begin{array}{lcl}
 \infty& \mbox{as}&  (\bfp/\Lambda)^2 \to 0,\\
     0 & \mbox{as}&  (\bfp/\Lambda)^2 \to \infty,
 \end{array}
\right.
\label{ir-func}
\end{equation}
but for definiteness we use the following IR cutoff function in our
analysis\footnote{In order to cutoff the momentum in an $M$-independent
way, we include $M$ in the cutoff function in this form.},
\begin{equation}
{R}^{(1)}_{\Lambda}(\bfp^2)=
\frac{\bfp^2}{2M}
\left[1-\exp\left[\left(\frac{\bfp^2}
{{\Lambda}^2}\right)^n\right]\right]^{-1},
\label{cutoffR}
\end{equation}
and take the sharp cutoff limit $n\to \infty$ in most cases. See the
next section.


For convenience we introduce the compact notation,
\begin{equation}
J_n\equiv(\eta,\eta^{\dagger}),
\quad U_n\equiv(N^{\dagger},N), \qquad (n=1,2),
\end{equation}
and write the generating functional ($\ref{generating}$) as
\begin{equation}
e^{i W_{\Lambda}[J]}= 
\int \mathcal{D}U\ 
\exp{i
\left\{S+\frac{1}{2}U_n\left(R_{\Lambda}\right)_{nm}U_m
+J_nU_n \right\}},
\label{generating2}
\end{equation}
where
\begin{equation}
(R_{\Lambda})_{nm}\equiv\left[
\begin{array}{ccc}
0                        & (R^{(1)}_{\Lambda})_{AB} \\
(R^{(2)}_{\Lambda})_{AB} & 0                        \\
\end{array}
\right],\qquad
(R^{(2)}_{\Lambda})_{AB}\equiv -(R^{(1)}_{\Lambda})_{BA}.
\end{equation}


The Legendre transform $\tilde{\Gamma}_{\Lambda}[\Phi]$ of
$W_\Lambda[J]$ may be defined in the standard way by introducing the
expectation value of $U_n$ in the presence of the source $J_n$,
\begin{equation}
\tilde{\Gamma}_{\Lambda}[\Phi]\equiv W_{\Lambda}[J]-J_n\Phi_n,
\qquad
\frac{\delta W_{\Lambda}}{\delta J_n}=\langle U_n\rangle\equiv\Phi_n,
\label{Legendre}
\end{equation}
where $\Phi_n\equiv(\psi^{\dagger},\psi)$ has been introduced.

It is more useful to define $\Gamma_{\Lambda}$, \textit{an averaged
action}, as follows,
\begin{equation}
\Gamma_{\Lambda}\equiv {\tilde\Gamma_{\Lambda}}-\frac{1}{2}\Phi_{n}
(R_{\Lambda})_{n m}\Phi_{m},
\end{equation}
which satisfies the following \textit{Legendre flow equation},
\begin{equation}
 \frac{d\Gamma_\Lambda}{d\Lambda}=\frac{i}{2}\Tr
  \left[
   \frac{dR_\Lambda}{d\Lambda}\left(\Gamma_{(2)}+R_\Lambda\right)^{-1}
  \right],
\end{equation}
where we have introduced the notation $\left( \Gamma_{(2)} \right)_{nm}
\equiv \delta^2 \Gamma_\Lambda/\delta \Phi_n \delta \Phi_m$ and $\Tr$
denotes the integration over momentum and also the trace in the internal
space. Although Legendre flow equation resembles a one-loop equation, it
contains nonperturbative information through a full propagator with IR
cutoff, $\left( \Gamma_{(2)} + R_{\Lambda} \right)^{-1}$.

We introduce a dimensionless parameter $t= \ln \left( \Lambda_0/\Lambda
\right)$ and $\tilde{\partial_{t}}$ which denotes the derivative with
respect to $t$ that acts \textit{only} on explicit $t$-dependence of the
IR cutoff function $R_{\Lambda}$.  The flow equation may be written in
an even simpler form,
\begin{equation}
\frac{d\Gamma_{\Lambda}}{dt}=\frac{i}{2}
\mbox{\rm Tr}\ \tilde{\partial}_t
\left[\ln\left(\Gamma_{(2)}+R_{\Lambda}\right)\right].
\label{floweq1}
\end{equation}

We may further simplify the expression by using the nonrelativistic
feature of the theory. For this purpose, let us split
$\left(\Gamma_{(2)} + R_{\Lambda}\right)$ into field independent part
$\mathcal{P}$ and field dependent part $\mathcal{F}$.  $\mathcal{P}$ is
the ``full propagator'', while $\mathcal{F}$ is the sum of the
multipoint ``vertices'' which have two internal lines and $\Phi_n$
attached to each of the external legs. The RHS of (\ref{floweq1}) can be
written as
\begin{equation}
\frac{i}{2} \mbox{\rm Tr}\ \tilde{\partial}_t
 \left[
  \ln{\mathcal{P}}+\left(\mathcal{P}^{-1}\mathcal{F}\right)
  -\frac{1}{2}\left(\mathcal{P}^{-1}\mathcal{F}\right)^2
  +\frac{1}{3}\left(\mathcal{P}^{-1}\mathcal{F}\right)^3+\cdots
  \right]. \label{derexp}
\end{equation}
In nonrelativistic theory, there are no anti-particles so that the loop
diagrams are very limited compared to relativistic theory; the diagrams
with anti-particle lines do not exist. The first term in
(\ref{derexp})  vanishes. (It is a field independent constant anyway.)
The second term also vanishes because there is no way to draw the loop
without an anti-particle line. The third and higher order terms contain
the non-vanishing diagrams.

In the following sections, we concentrate on the two-nucleon sector, to
which only the four-nucleon (4N) operators contribute. It is easy to see
that only the third term contains such contributions. (It also contains
the contributions to the other sectors.)  The other sectors get
contributions from several terms in (\ref{derexp}).  To the
three-nucleon sector, for example, both the third and the fourth terms
contain the contributions.

In this way, we end up with the following reduced equation for the 4N
operators,
\begin{equation}
 \left.
  \frac{d\Gamma_{\Lambda}}{dt}
 \right|_{\mbox{\scriptsize 4N}}
=-\left.\frac{i}{4} 
   \mbox{\rm Tr}\ \tilde{\partial_{t}}
   \left[({\mathcal P}^{-1}{\mathcal F})^{2}\right]
  \right|_{\mbox{\scriptsize 4N}}.
\label{floweq2}
\end{equation}


Although the above equation is \textit{exact}, one needs an
approximation to solve it. We consider a simple truncation of the space
of operators. Namely, we consider only the operators of leading orders
in derivative expansion. The approximation is based on our hope that,
even though some operators get large anomalous dimensions, their
``ordering'' of importance would not change; the lower the canonical
dimension is, the lower the scaling dimension would be. Of course some
mixing of operators would occur and should be taken into account
properly. There is no absolute justification for this hope. One should
examine its validity by actually enlarging the space of operators and
confirming the stability of the results against the enlargement.


An actual calculation goes as follows. We substitute the ``ansatz'' for
the averaged action consisting of the operators up to a certain order
into the above flow equation. The ``ansatz'' determines the explicit
forms of $\mathcal{P}$ and $\mathcal{F}$. We then compare the
coefficients of the operators on both sides. Higher oder operators which
emerge in the RHS of (\ref{floweq2}) are disregarded (``projected
out''). The LHS contains the derivatives of the couplings while the RHS
does not, thus the comparison leads to a set of first-order differential
equations for the couplings, the RG equations.


It is important to note that a WRG transformation generates all kinds of
operators allowed by the symmetry of the theory including the so-called
``redundant'' operators, the operators which may be eliminated by the
use of equations of motion. The ``use of equations of motion'' actually
means a field redefinition, which eliminates the ``redundant'' term. In
the previous paper\cite{Harada:2005tw}, we showed that the field
redefinition gives rise to a nontrivial Jacobian, and that ``redundant''
operators play an important role in a WRG analysis.

\section{RG analysis of pionless NEFT}
\label{sec:pionless}

\subsection{pionless NEFT}


In this section, we consider the RG flow of the pionless NEFT, in which
only nonrelativistic nucleons are treated as explicit degrees of
freedom.  Effects of anti-nucleons, pions, and other heavier mesons and
baryons are integrated out and hidden in the values of the coupling
constants.  Thus the theory is expected to describe the interactions of
nonrelativistic nucleons with external momenta sufficiently smaller than
pion mass. Although our approach is general and not restricted to this
theory, it serves as the simplest example which illustrates the
essential points.


In a WRG analysis, symmetry is extremely important. We assume Galilean
invariance, rotational symmetry, and invariance under charge
conjugation, parity, and time-reversal. Furthermore, we assume exact
isospin symmetry, and ignore electromagnetic and weak interactions. In
the following, we will focus only on two channels, ${}^1S_0$ and
${}^3S_1$--${}^3D_1$, in the two-nucleon sector. The extension to higher
partial waves is easy.


We study the RG flow for this theory by using Legendre flow equation.
We consider the following ``ansatz'' for the averaged action, keeping
only the operators with (canonical) dimension up to eight. 
\begin{eqnarray}
\Gamma_{\Lambda}&=&\int d^4x
 \bigg[
 N^{\dagger}\left(i\partial_t+\frac{{\nabla}^2}{2M}\right)N
 \nonumber \\
 &&\left\{
  \begin{array}{lcl}
   -C^{(S)}_0\mathcal{O}^{(S)}_0
    +C^{(S)}_2\mathcal{O}^{(S)}_2
    +2B^{(S)}\mathcal{O}^{(SB)}_2\bigg],
    &\quad& \mbox{(${}^1S_0$\ channel)}\\
   -C^{(T)}_0\mathcal{O}^{(T)}_0
    +C^{(T)}_2\mathcal{O}^{(T)}_2
    +2B^{(T)}\mathcal{O}^{(TB)}_2
    +C^{(SD)}_2\mathcal{O}^{(SD)}_2\bigg],
    &\quad& \mbox{(${}^3S_1$--${}^3D_1$\ channel)}\\
  \end{array}
\right.
\label{truncation}
\end{eqnarray}
where the operators in the ${}^1S_0$ are given by
\begin{subequations}
\begin{eqnarray}
 \mathcal{O}^{(S)}_0&=&
  \left(
   N^TP_a^{(^1 S_0)}N
  \right)^{\dagger}
  \left(
   N^{T}P_a^{(^1 S_0)}N
  \right), \\
 \mathcal{O}^{(S)}_2&=&
 \left[
  \left(
   N^TP_a^{(^{1}S_0)}N
  \right)^{\dagger}
  \left(
   N^{T}P_a^{(^{1}S_0)}\overleftrightarrow{\nabla}^2N
  \right)+ h.c.
 \right], \\
\mathcal{O}^{(SB)}_2&=&
\left[
 \left\{
  N^TP_a^{(^{1}S_0)}\left(i\partial_t+\frac{\nabla^2}{2M}\right)N
 \right\}^{\dagger}
 \left(
  N^TP_a^{(^{1}S_0)}N
 \right)+h.c.
\right], 
\end{eqnarray}
\end{subequations}
and in the ${}^3S_1$--${}^3D_1$ channel, 
\begin{subequations}
\begin{eqnarray}
 \mathcal{O}^{(T)}_0&=&
  \left(
   N^TP_i^{(^{3}S_1)}N
  \right)^{\dagger}
  \left(
   N^{T}P_i^{(^3S_1)}N
  \right),\\
 \mathcal{O}^{(T)}_2&=&
  \left[
   \left(
    N^TP_i^{(^{3}S_1)}N
   \right)^{\dagger}
   \left(
    N^{T}P_i^{(^3S_1)}\overleftrightarrow{\nabla}^2N
   \right)+ h.c.
  \right], \\
 \mathcal{O}^{(SD)}_2&=&
  \left[
   \left(
    N^TP_i^{(^{3}S_1)}N
   \right)^{\dagger}
   \left\{
    N^{T}
    \left(
     \overleftrightarrow{\nabla}_i\overleftrightarrow{\nabla}_j
     -\frac{1}{3}\delta_{ij}\overleftrightarrow{\nabla}^2
     \right)
    P^{(^{3}S_1)}_{j}N
   \right\}+ h.c.
   \right], \\
 \mathcal{O}^{(TB)}_2&=&
  \left[
   \left\{
    N^TP_i^{(^{3}S_1)}\left(i\partial_t+\frac{\nabla^2}{2M}\right)N
   \right\}^{\dagger}
   \left(
    N^TP_i^{(^{3}S_1)}N\right)+h.c.
  \right],\\
\end{eqnarray}
\end{subequations}
where we have introduced the notation $\overleftrightarrow{\nabla}^2
\equiv \overleftarrow{\nabla^2} + \overrightarrow{\nabla^2} -
2\overleftarrow{\nabla}\cdot\overrightarrow{\nabla}$ and the projection
operators\cite{Fleming:1999ee},
\begin{equation}
 P_a^{({}^{1}S_0)}\equiv\frac{1}{\sqrt{8}}\sigma^2\tau^2\tau^a,
  \qquad
 P_k^{({}^{3}S_1)}\equiv\frac{1}{\sqrt{8}}\sigma^2\sigma^k\tau^2,
\end{equation}
for the ${}^1S_0$ channel and the ${}^3S_1$ channel respectively. The
nucleon field $N(x)$ with mass $M$ is an isospin doublet nonrelativistic
two-component spinor. Pauli matrices $\sigma^i$ and $\tau^a$ act on spin
indices and isospin indices respectively. The two channels are
completely decoupled, and thus we can consider each channel separately.


Note that we have not included the wave function renormalization because
there is no such contribution due to the nonrelativistic feature. We
neither include six-nucleon operators and higher because of the nucleon
number conservation. The more-than-two-nucleon sectors are completely
decoupled from the two-nucleon sector which we are interested in.


Note also that we have included redundant operators,
$\mathcal{O}^{(SB)}_2$ and $\mathcal{O}^{(TB)}_2$, because they are
necessary in a consistent WRG analysis.


 \begin{figure}
 \includegraphics[width=12cm,clip]{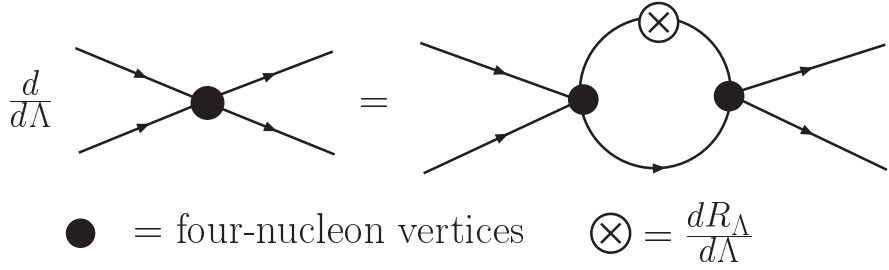} 
  \caption{\label{loop}The loop diagram which contributes to the Legendre
  flow equation for four-nucleon operators.}
 \end{figure}

Let us consider the simplest example of the loop calculation in order to
explain how we maintain the Galilean invariance. The diagrams which
contribute to the Legendre flow equation are of the type shown in
FIG.~\ref{loop}, with various factors at the vertices.  The propagator
is IR cutoff by the function (\ref{cutoffR}), so that we are interested
in the following integral,
\begin{equation}
 I\equiv \int \frac{d^4k}{(2\pi)^4}
  \frac{i}{
   P^0/2+k^0
   -\mathcal{R}\left(\left(\bfP/2+\bfk\right)^2\right)
   +i\epsilon
   }\cdot
   \frac{i}{
   P^0/2-k^0
   -\mathcal{R}\left(\left(\bfP/2-\bfk\right)^2\right)
   +i\epsilon
   }\ ,
\end{equation}
where we have introduced 
\begin{equation}
 \mathcal{R}\left(\bfk^2\right)=
  \frac{\bfk^2}{2M}-R^{(1)}_{\Lambda}\left(\bfk^2\right),
\end{equation}
and $(P^0, \bfP)$ is the momentum flowing in (and out of) the
diagram. Note that we assign the loop momentum $(k^0,\bfk)$ so that $I$
is invariant under $\bfP \rightarrow -\bfP$. Note also that there may be
additional momentum factors from the vertices but they do not affect the
following argument. After integrating $k^0$, we have
\begin{equation}
 I=\int \frac{d^3k}{(2\pi)^3}
  \frac{i}{
  P^0
  -\mathcal{R}\left(\left(\bfP/2-\bfk\right)^2\right)
  -\mathcal{R}\left(\left(\bfP/2+\bfk\right)^2\right)
  +i\epsilon
  }. \label{twoR}
\end{equation}
It is important to note that, if the IR cutoff were absent, the
propagator would be
\begin{equation}
 \frac{i}{
  P^0-\frac{\bfP^2}{4M} -\frac{\bfk^2}{M}+i\epsilon
  },
\end{equation}
where the combination $\bfP^2/4M$ describes the center-of-mass kinetic
energy of the two nucleons while $\bfk^2/M$ represents the kinetic
energy of the relative motion. It is now clear that \textit{the only
way} to maintain the Galilean invariance is to cutoff the relative
momentum. It is therefore physically legitimate to replace (\ref{twoR})
with
\begin{equation}
 \int \frac{d^3k}{(2\pi)^3}
  \frac{i}{
  P^0-\frac{\bfP^2}{4M}-\frac{\bfk^2}{M}
  +2R_{\Lambda}^{(1)}\left(\bfk^2\right)
  +i\epsilon
  }. \label{oneR}
\end{equation}


The integral may be evaluated with an arbitrary parameter $n$. A useful
formula for a typical loop integral is given in Appendix
\ref{sec:formulae}. In the following, we consider only the sharp cutoff
limit $n \rightarrow \infty$, for which the RG equation becomes the
simplest. We also look at the dependence of the fixed points and the
anomalous dimensions against the variation of $n$. They are shown in
Appendix \ref{sec:n-dep}.

\subsection{RG flows and fixed points}


It is useful to define the dimensionless variables as follows and write
the RG equation in terms of these variables. In the ${}^1S_0$ channel, we
introduce\footnote{The factor $M$ comes from the scale transformation
property in nonrelativistic theory, $t'=\lambda^2 t$, $\bfx'=\lambda
\bfx$, where $\lambda$ is the scale factor. Under this transformation,
the nucleon field transforms as $N(t,
\bfx)=\lambda^{\frac{3}{2}}N'(t',\bfx')$, and thus we have
$\mathcal{O}^{(S)}_0 = \lambda^6\mathcal{O'}^{(S)}_0$ for example. The
$d^4x$ gives the factor $\lambda^{-5}$, and the additional factor
$\lambda^{-1}$ is supplied by the $\Lambda$-dependence of the coupling
$C_0^{(S)}\propto \Lambda^{-1}$ to cancel the $\lambda^6$. To have the
right mass dimension, $C_0^{(S)}$ has the $1/M$ dependence. Similar
consideration leads to the correct $\Lambda$ and $M$ dependence of other
couplings.},
\begin{equation}
x\equiv \frac{M\Lambda}{2\pi^2} C_0^{(S)},\quad
y\equiv \frac{M\Lambda^3}{2\pi^2} 4C_2^{(S)},\quad
z\equiv \frac{\Lambda^3}{2\pi^2} B^{(S)},
\label{dimensionlesssinglet}
\end{equation}
in terms of which the RG equations are given by
\begin{subequations}
\label{RGEsinglet}
\begin{eqnarray}
 \frac{dx}{dt}&=&
  -x
  -\Biggl[
  x^2+2xy+y^2+2xz+2yz+z^2
  \Biggr], \\
 \frac{dy}{dt}&=&
  -3y
  -\Biggl[
  \frac{1}{2}x^2+2xy+\frac{3}{2}y^2+yz-\frac{1}{2}z^2
  \Biggr], \\
 \frac{dz}{dt}&=&
  -3z
  +\Biggl[
  \frac{1}{2}x^2+xy+\half y^2-xz-yz-\frac{3}{2}z^2
  \Biggr].
\end{eqnarray}
\end{subequations}
Note that the RG equations are quadratic in four-nucleon
couplings due to the nonexistence of anti-nucleons in our
nonrelativistic formulation,
up to the first terms, which come from the canonical scaling.

Similarly in the ${}^3S_1$--${}^3D_1$ channel, we introduce
\begin{eqnarray}
x'\equiv \frac{M\Lambda}{2\pi^2} C_0^{(T)}, \quad
y'\equiv \frac{M\Lambda^3}{2\pi^2}4 C_2^{(T)}, \quad 
z'\equiv \frac{\Lambda^3}{2\pi^2} B^{(T)}, \quad
w'\equiv \frac{M\Lambda^3}{2\pi^2}\frac{4}{3}C_2^{(SD)},
\label{dimensionlesstriplet}
\end{eqnarray}
and we have the following RG equations,
\begin{subequations}
\label{RGEtriplet}
\begin{eqnarray}
 \frac{dx'}{dt}&=&
  -x'
  -\Biggl[
  {x'}^{2}+2x'y'+{y'}^{2}+2x'z'+2y'z'+{z'}^{2}+2{w'}^{2}
  \Biggr],\\
 \frac{dy'}{dt}&=&
 -3y'
 -\Biggl[
 \frac{1}{2}{x'}^{2}+2x'y'+\frac{3}{2}{y'}^{2}+y'z'-\frac{1}{2}{z'}^{2}
 +{w'}^{2}
 \Biggr],\\
 \frac{dz'}{dt}&=&
 -3z'
 +\Biggl[
 \frac{1}{2}{x'}^{2}+x'y'+\half {y'}^{2}-x'z'-y'z'-\frac{3}{2}{z'}^{2}
 +{w'}^{2}
 \Biggr],\\
 \frac{dw'}{dt}&=&
 -3w'
 -\Biggl[
 x'w'+y'w'+z'w'
 \Biggr].
\end{eqnarray} 
\end{subequations}
Note that the RG equations (\ref{RGEtriplet}) for the
${}^3S_1$--${}^3D_1$ channel are identical to the RG equations
(\ref{RGEsinglet}) for the ${}^1S_0$ channel if $w'$ is set equal to
zero. (This is actually a solution.)

\begin{figure}[htbp]
\begin{tabular}{cc}
\begin{minipage}[t]{0.45\linewidth}
\begin{flushleft}
\includegraphics[width=\linewidth,clip]{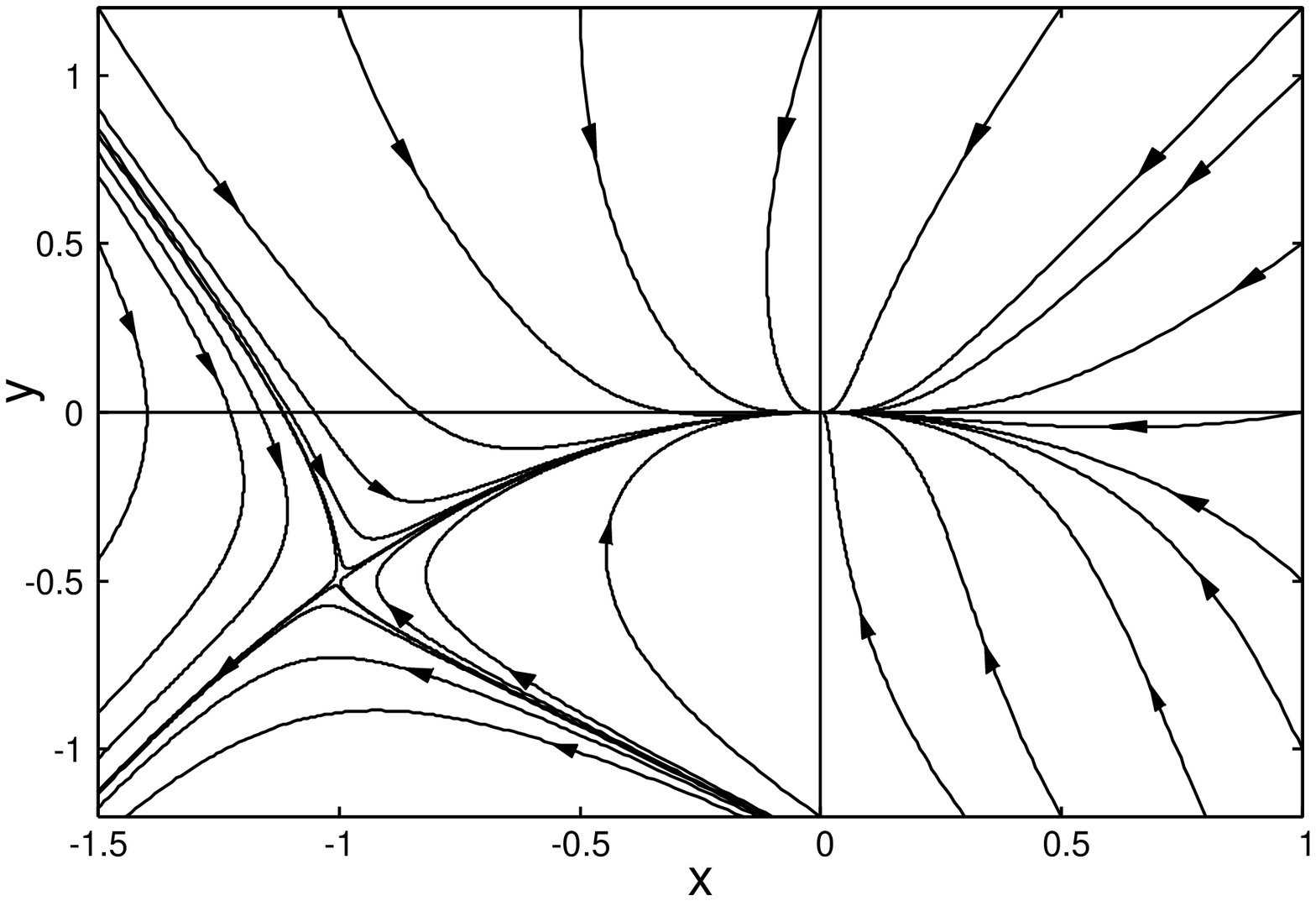}
\caption{\label{singlet}The RG flow for the ${}^1S_0$ channel
 projected onto the $C_0^{(S)}$--$C_2^{(S)}$ plane. Flow lines
 start with the arbitrarily chosen value $z=0$ at the edges of the
 graph.}
\end{flushleft}
\end{minipage}\quad\quad\quad
\begin{minipage}[t]{0.45\linewidth}
\begin{flushright}
\includegraphics[width=\linewidth,clip]{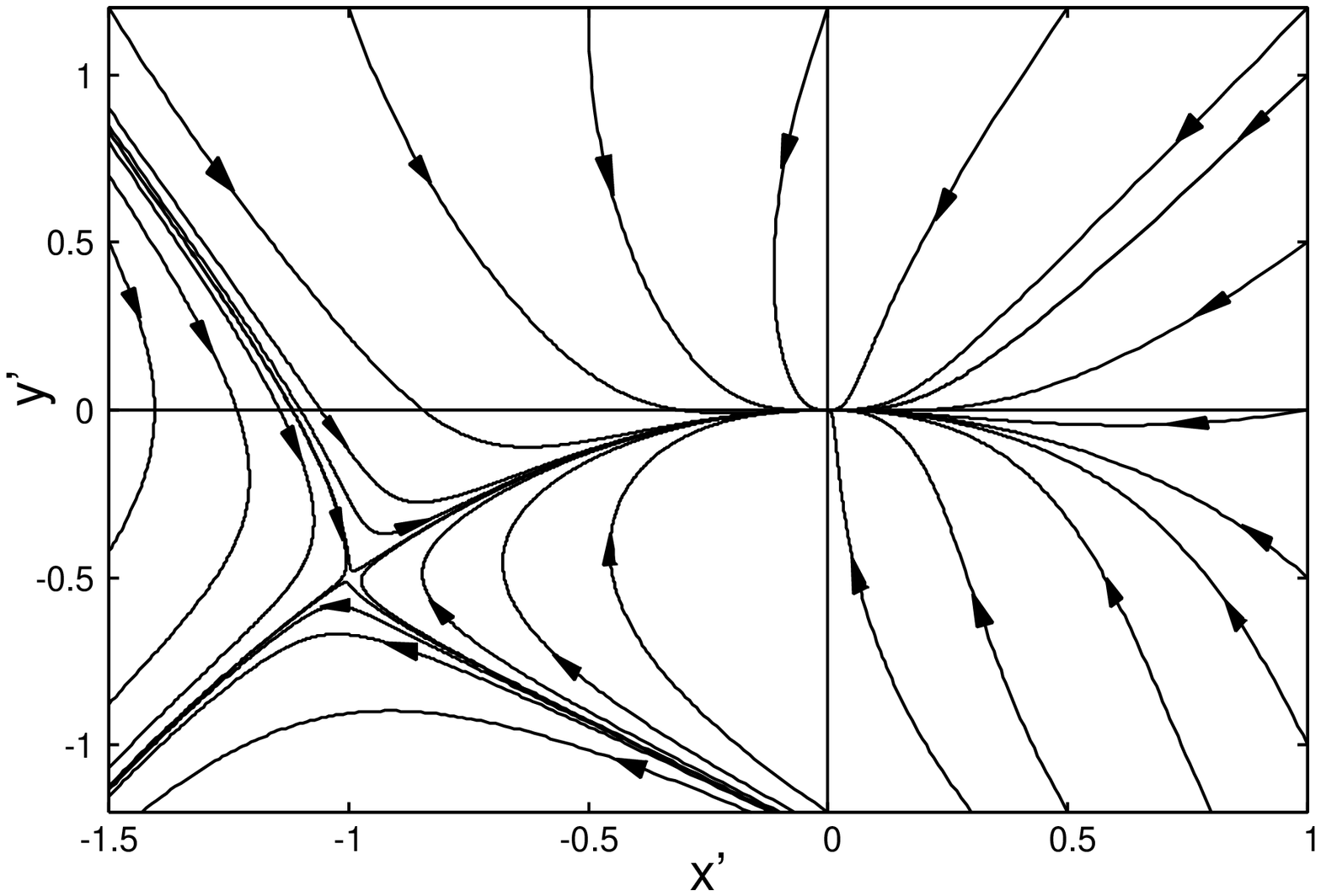}
\caption{\label{triplet}The RG flow for the ${}^3S_1$--${}^3D_1$
 channel projected onto the $C_0^{(T)}$--$C_2^{(T)}$ plane. Flow lines
 start with the arbitrarily chosen values $z'=0$ and $w'=0.1$ at the
 edges of the graph.}
\end{flushright}
\end{minipage}
\end{tabular}
\end{figure}

We can now draw the RG flow diagram.  FIG.~\ref{singlet} shows the RG
flow for the ${}^1S_0$ channel and FIG.~\ref{triplet} for the
${}^3S_1$--${}^3D_1$ channel up to operators with dimension eight. Both
flows are projected on to the two-dimensional plane spanned by the
lowest order coupling constants.
As expected from the RG equations (\ref{RGEsinglet}) and
(\ref{RGEtriplet}), the RG flows for the ${}^1S_0$ channel and for the
${}^3S_1$--${}^3D_1$ channel have similar structure.  First of all,
there are two fixed points in both RG flows; the trivial IR fixed point
and the non-trivial UV fixed point.  Secondly, both RG flows have
\textit{two phases}; one in which all points approaches the trivial
fixed point in the limit of vanishing cutoff, and the other one in which
all points approaches infinity at certain finite cutoff scale. The
nontrivial fixed points are on the phase boundaries.


Location of the fixed points are given by solutions of the equations,
$dx/dt = dy/dt = dz/dt = 0$, for the ${}^1S_0$ channel, and $dx'/dt =
dy'/dt = dz'/dt = dw'/dt = 0$ for the ${}^3S_1$--${}^3D_1$ channel.  As
explained in Appendix~\ref{sec:n-dep}, we find four fixed points in each
channel for an arbitrary value of the parameter $n$. One of them is
extremely unstable against the variation of $n$, and may be considered as
``spurious'', i.e., it is an artifact of our truncation. The rest are
stable and take the following values in the $n\rightarrow \infty$ limit,

\underline{${}^1S_0$ channel}:

\begin{equation}
\left(x^{\star},y^{\star},z^{\star}\right)=
\left(0,0,0\right),\quad
\left(-1,-\frac{1}{2},\frac{1}{2}\right),\quad
\left(-9,\frac{15}{2},-\frac{3}{2}\right)
\end{equation}

\underline{${}^3S_1$--${}^3D_1$ channel}:

\begin{equation}
\left({x'}^{\star},{y'}^{\star},{z'}^{\star},{w'}^{\star}\right)=
\left(0,0,0,0\right),\quad
\left(-1,-\frac{1}{2},\frac{1}{2},0\right),\quad
\left(-9,\frac{15}{2},-\frac{3}{2},0\right).
\end{equation}

We also disregard the third fixed points in both channels, because they
are irrelevant to the discussion of the power counting. The
eigenoperators at these fixed points are complex and have complex
eigenvalues (scaling dimensions). It means that the operators do not
have the definite scaling property there.

We obtain the scattering amplitudes explicitly in both channels in
Appendix~\ref{sec:amplitudes} as we did in the previous
paper\cite{Harada:2005tw}, and find that the RG equations, which look
much more complicated than (\ref{RGEsinglet}) and (\ref{RGEtriplet}),
have only two fixed points in each channel and they exactly agree with
the first two fixed points obtained here\footnote{Actually, the RG
equation obtained in the previous paper\cite{Harada:2005tw} is those for
the ${}^1S_0$ channel.}. 

\subsection{scaling dimensions}

We now study the behavior of the coupling constants around each fixed
point by considering the linearized RG equations. 


Let us first consider the trivial fixed points. In the ${}^1S_0$
channel, the trivial fixed point is at $(x^\star, y^\star, z^\star) =
(0,0,0)$. We consider a small deviation $(\delta x, \delta y, \delta z)$
around it and linearize the RG equations (\ref{RGEsinglet}),
\begin{equation}
\frac{d}{dt}
 \left(
  \begin{array}{c}
   \delta x\\
   \delta y\\
   \delta z
  \end{array}
 \right)
 =\left(
   \begin{array}{ccc}
    -1&0&0\\
    0&-3&0\\
    0&0&-3
   \end{array}
  \right)
 \left(
  \begin{array}{c}
   \delta x\\
   \delta y\\
   \delta z\\
  \end{array}
 \right).
\end{equation}
The eigenvalues and corresponding eigenvectors are trivially obtained,
\begin{equation}
 \nu_1=-1:\ 
 u_1=\left(
      \begin{array}{c}
       1\\
       0\\
       0
      \end{array}
      \right),\quad
 \nu_2=-3:\ 
 u_2=\left(
      \begin{array}{c}
       0\\
       1\\
       0
      \end{array}
     \right),\quad
 \nu_3=-3:\ 
 u_3=\left(
      \begin{array}{c}
       0\\
       0\\
       1
      \end{array}
     \right).
\end{equation}
Namely, the couplings $C^{(S)}_0$, $C^{(S)}_2$, and
$B^{(S)}$ have the  scaling dimension $-1$, $-3$, and $-3$
respectively. This is what one naively expects from their canonical
scaling properties.

Similarly in the ${}^3S_1$--${}^3D_1$ channel, we consider a deviation
$(\delta x', \delta y', \delta z', \delta w')$ around the trivial fixed
point $({x'}^\star, {y'}^\star, {z'}^\star, {w'}^\star) = (0,0,0,0)$,
and find the following linearized RG equations from
Eq.~(\ref{RGEtriplet}),
\begin{equation}
 \frac{d}{dt}
  \left(
   \begin{array}{c}
    \delta x'\\
    \delta y'\\
    \delta z'\\
    \delta w'
   \end{array}
  \right)
 =\left(
   \begin{array}{cccc}
    -1&0&0&0\\
    0&-3&0&0\\
    0&0&-3&0\\
    0&0&0&-3
   \end{array}
  \right)
 \left(
  \begin{array}{c}
   \delta x'\\
   \delta y'\\
   \delta z'\\
   \delta w'
  \end{array}
 \right),
\end{equation}
with the eigenvalues and corresponding eigenvectors,
\begin{eqnarray}
 \nu_1&=&-1:\ 
 u_1=\left(
      \begin{array}{c}
       1\\
       0\\
       0\\
       0
      \end{array}
      \right),\quad
 \nu_2=-3:\ 
 u_2=\left(
      \begin{array}{c}
       0\\
       1\\
       0\\
       0
      \end{array}
      \right),\quad
 \nu_3=-3:\ 
 u_3=\left(
      \begin{array}{c}
       0\\
       0\\
       1\\
       0
      \end{array}
     \right),\nonumber \\
 \nu_4&=&-3:\ 
 u_4=\left(
      \begin{array}{c}
       0\\
       0\\
       0\\
       1
      \end{array}
     \right).
\end{eqnarray}
The coupling constants $C^{(T)}_0$, $C^{(T)}_2$, $B^{(T)}$, and
$C^{(SD)}_2$ have the scaling dimension $-1$, $-3$, $-3$, and $-3$
respectively.


Note that all of the scaling dimensions are negative at the trivial
fixed point and thus the corresponding operators are irrelevant, i.e.,
they should be treated perturbatively.


At the nontrivial fixed point, on the other hand, something much more
interesting happens. In the ${}^1S_0$ channel, the deviations from the
nontrivial fixed point $(x^\star, y^\star, z^\star) = (-1, -1/2, 1/2)$
satisfy the following linearized RG equations,
\begin{equation}
\frac{d}{dt}
 \left(
  \begin{array}{c}
   \delta x\\
   \delta y\\
   \delta z
  \end{array}
 \right)
 =\left(
   \begin{array}{ccc}
    1 & 2& 2\\
    2 & 0& 1\\
    -2&-2&-3
   \end{array}
  \right)
 \left(
  \begin{array}{c}
   \delta x\\
   \delta y\\
   \delta z\\
  \end{array}
 \right),
\end{equation}
with the eigenvalues and corresponding eigenvectors,
\begin{equation}
 \nu_1=+1:\ 
 u_1=\left(
      \begin{array}{c}
       1\\
       1\\
       -1
      \end{array}
      \right),\quad
 \nu_2=-1:\ 
 u_2=\left(
      \begin{array}{c}
       0\\
       -1\\
       1
      \end{array}
     \right),\quad
 \nu_3=-2:\ 
 u_3=\left(
      \begin{array}{c}
       2\\
       -1\\
       -2
      \end{array}
     \right).
\end{equation}
Note that the scaling dimensions change drastically. Note also that the
eigenoperators are not monomials but linear combinations of the
operators. This is a typical feature of the momentum cutoff
regularization.

Similar results are obtained in the ${}^3S_1$--${}^3D_1$ channel, where
the linearized equations around the nontrivial fixed point $({x'}^\star,
{y'}^\star, {z'}^\star, {w'}^\star)=(-1,-1/2,1/2,0)$ become
\begin{equation}
 \frac{d}{dt}
  \left(
   \begin{array}{c}
    \delta x'\\
    \delta y'\\
    \delta z'\\
    \delta w'
   \end{array}
  \right)
  =\left(
    \begin{array}{cccc}
     1 & 2& 2&0\\
     2 & 0& 1&0\\
     -2&-2&-3&0\\
     0 & 0& 0&-2
    \end{array}
   \right)
  \left(
   \begin{array}{c}
    \delta x'\\
    \delta y'\\
    \delta z'\\
    \delta w'
   \end{array}
  \right),
\end{equation}
giving the following eigenvalues and corresponding eigenvectors,
\begin{eqnarray}
 \nu_1&=&+1:\ 
  u_1=\left(
       \begin{array}{c}
	1\\
	1\\
	-1\\
	0
       \end{array}
      \right),\quad
  \nu_2=-1:\ 
  u_2=\left(
       \begin{array}{c}
	0\\
	-1\\
	1\\
	0
       \end{array}
      \right),\quad
  \nu_3=-2:\ 
  u_3=\left(
       \begin{array}{c}
	2\\
	-1\\
	-2\\
	0
       \end{array}
      \right),\nonumber \\
 \nu_4&=&-2:\ 
  u_4=\left(
       \begin{array}{c}
	0\\
	0\\
	0\\
	1
       \end{array}
      \right).
\end{eqnarray}
Note that the structure is essentially the same as that in the ${}^1S_0$
channel, with $\delta w'$ decoupled.


The most important result at the nontrivial fixed point is that
\textit{there is a relevant coupling} in each channel. The coupling
$u_1$ has dimension $+1$, and according to the argument given in
Sec.~\ref{sec:powercounting}, the corresponding operator should be
treated nonperturbatively, while the others perturbatively. The
operators should be ordered according to the scaling dimensions. The
less the dimension of a coupling constant is, the less the importance of
the corresponding operator is.


As we have done in the previous paper\cite{Harada:2005tw}, RG equations
may be obtained by directly calculating the amplitudes and requiring
them to be independent of the ``floating'' scale $\Lambda$. See
Appendix~\ref{sec:amplitudes} for the results. The resulting RG
equations are much more complicated. It is interesting, however, the
location of the fixed points and the scaling dimensions are precisely
the same as those obtained here with the Legendre flow equation. The
eigenvectors are a bit different, though.

\subsection{physical meaning of the RG phases}

Let us consider the real world and examine the physical meaning of what
we have found in the two-nucleon system.


As mentioned in Introduction, the two-nucleon system is considered as
``fine-tuned,'' in the sense that the scattering length is unnaturally
large. It suggests that the actual system is near the nontrivial fixed
point in each channel. The closer to the fixed point it is, the longer
the scattering length is. 


What then makes the physical difference between the ${}^1S_0$ and the
${}^3S_1$--${}^3D_1$ channels, both of which are close to the nontrivial
fixed points? The answer to this question is provided by the flow
diagram; the nontrivial fixed point is \textit{on the boundary} between
two phases. In one of them, the flow goes toward the trivial fixed
point in the IR. We therefore call this \textit{the weak coupling
phase}. In the other, the flow goes toward stronger couplings, thus
we call it \textit{the strong coupling phase}. It is natural to think
that the real two-nucleon system in the ${}^1S_0$ channel, in which there
is no bound state, is in the weak coupling phase, while that in the
${}^3S_1$--${}^3D_1$ channel, in which there is a shallow bound state,
deuteron, is in the strong coupling phase.


In order to make the physical picture more transparent, it is useful to
examine the four-nucleon (two-nucleon scattering) amplitudes.  As shown
in Ref.~\cite{Harada:2005tw} and in Appendix~\ref{sec:amplitudes}, the
four-nucleon amplitudes may be obtained explicitly in the derivative
expansion.  By substituting the solution of the linearized RG equation
in the ${}^1S_0$ channel obtained from the direct method,
\begin{eqnarray}
 \left(
  \begin{array}{c}
   \delta x\\
   \delta y\\
   \delta z
  \end{array}
 \right)
 =a
 \left(
  \begin{array}{c}
   2\\
   1\\
   -4
  \end{array}
 \right)\left(\frac{\Lambda}{\Lambda_0}\right)^2
 +b
 \left(
  \begin{array}{c}
   0\\
   -1\\
   1
  \end{array}
 \right)\left(\frac{\Lambda}{\Lambda_0}\right)
 +c
 \left(
  \begin{array}{c}
   1\\
   1\\
   -1
  \end{array}
 \right)\left(\frac{\Lambda_0}{\Lambda}\right), \label{singletvariation}
\end{eqnarray}
where $a$, $b$, and $c$ are small dimensionless constants, into the
amplitude in the center-of-mass frame, we obtained the renormalized
(off-shell) amplitude near the nontrivial fixed
point\cite{Harada:2005tw},
\begin{equation}
 \left.\mathcal{A}^{-1}(p^0, {\bfp'}^2, \bfp^2)\right|_*
  =-\frac{M}{4\pi}
  \left[
   \frac{2c}{\pi}\Lambda_0
   -\frac{4b}{\pi}\left(\frac{Mp^0}{\Lambda_0}\right)
   -i\sqrt{M p^0}
  \right]+\cdots.
\end{equation}
where ellipsis denotes higher orders in $1/\Lambda_0$.
 

By comparing it with the effective range expansion,
\begin{equation}
 \left.\mathcal{A}^{-1}\right|_{on-shell}
  = -\frac{M}{4\pi}
  \left[
   -\frac{1}{\alpha}
   +\frac{1}{2}r p^2
   -i p
  \right]+\cdots,
\end{equation}
on the mass shell $p = \sqrt{Mp^0} = \left|\bfp\right| =
\left|\bfp'\right|$, one sees the scattering length $\alpha$ and the
effective range $r$ are given by
\begin{equation}
 \alpha=-\frac{\pi}{2c\Lambda_0},\quad r=-\frac{8b}{\pi\Lambda_0}.
\end{equation}
That is, \textit{the inverse of the scattering length is identified as
the relevant coupling}. The fine-tuning $\left|\alpha\right| \gg
\Lambda_0^{-1}$ indeed corresponds to $\left|c\right| \ll 1$. It is also
important that the sign of the coupling $c$ distinguishes the phases,
namely, \textit{the inverse of the scattering length is the order
parameter}.  It vanishes on the boundary, the critical surface. Because
the effective range is of natural size, $r\sim \Lambda_0^{-1}$, it is
more accurate to say that the real two-nucleon system is close to the
critical surface.


The location of the pole which is found within the range of EFT is given
by
\begin{eqnarray}
p_0\simeq-\frac{\Lambda_0^2}{M}\left(\frac{2c}{\pi}\right)^2
 \left[1+\frac{16bc}{\pi^2}\right].
\end{eqnarray}
A similar expression is obtained for the pole in the
${}^3S_1$--${}^3D_1$ channel.  The pole on the physical Riemann sheet
that corresponds to a bound state occurs only when $c<0$, i.e., in the
strong coupling phase, while there is no restriction on $b$.

\section{Comments on other power counting schemes}
\label{sec:comments}


As we emphasized in Sec.~\ref{sec:powercounting}, our way of determining
the power counting on the basis of the scaling dimensions is general and
nonperturbative. It does not depend on any specific
regularization/renormalization scheme, nor on which fixed point we are
looking at.

In this section, we would like to comment on other works from our point
of view and make the connection between their power counting and
ours. It leads us to a deeper understanding of the power counting issues.


First of all, we would like to mention the work done by Birse, McGovern,
and Richardson\cite{Birse:1998dk}, which is very close to ours in
philosophy. They consider the most general potential and the Wilsonian
RG equation satisfied by it. The ``fixed point'' potential depends only
on the energy,
\begin{equation}
 V_0(p,\Lambda)=-\frac{2\pi^2}{M}
  \left[
   \Lambda-\frac{p}{2}\ln\frac{\Lambda+p}{\Lambda-p}
  \right]^{-1},
  \label{birsepot}
\end{equation}
where $p=\sqrt{ME}$, with the $E$ being the center-of-mass energy. They
then consider the perturbation around the ``fixed point'' potential, and
find the eigenvalues of the linearized RG equations. The first few
eigenvalues coincide with ours. They use the information to determine
the power counting. It is important to note that the ``energy-dependent
perturbations'' have the spectrum $\nu=-1, 1, 3, \cdots$, while the
``momentum-dependent perturbations'' have $\nu=2,4,5,\cdots$. (Note that
their $\nu$'s have opposite signs to the scaling dimensions used in this
paper.)


To compare their results with ours, it is useful to translate our
operator formulation to the potential one. Our operators may be
equivalent to the potential,
\begin{equation}
 V(p^0, \bfp, \bfp')=C_0+4C_2(\bfp^2+{\bfp'}^2)
  -2B\left(p^0-\frac{\bfp^2+{\bfp'}^2}{2M}\right)+\cdots,
\end{equation}
where we suppress the superscripts which denote the channel. If we
substitute the fixed point values of the couplings, we have
\begin{equation}
 V^\star(p^0, \bfp, \bfp')=-\frac{2\pi^2}{M\Lambda}
  \left[
   1+\frac{Mp^0}{\Lambda^2}
  \right]+\cdots,
\end{equation}
which is exactly the expansion of the potential (\ref{birsepot}), with
$p^0$ replaced by $E$. Note that the momentum dependence cancels at the
fixed point. This may be an evidence that both approaches are
essentially the same. Actually, the ``perturbations'' with the scaling
dimensions $\nu_1=1$ and $\nu_2=-1$ corresponds to the potentials that
depends only on $p^0$, while that with $\nu_3=-2$ to a
momentum-dependent one, in agreement with their findings\footnote{A
careful study reveals that the eigenvector for $\nu_3=-2$ does not
agree with theirs, though those for $\nu_1=1$ and $\nu_2=-1$ do. The
reason seems that the reaction matrix they consider does not have a
direct connection to the scattering amplitude which we work with
\textit{off the mass shell}.}. 


For this simplest pionless NEFT in the two-nucleon sector, the method
based on the potential has an advantage.  On the other hand, because our
approach is based on the general principle of fully-fledged quantum
field theory, we could treat more complicated systems in a unified and
systematic way. For example, we could treat three-nucleon problems
without worrying about the off-shell ambiguities, and fully relativistic
systems for which the potential picture is not valid. The important
point is that our work gives a field-theoretical foundation for their
approach and has opened a window to wider applications.

In the following, we will make comments on other power counting schemes
on the basis of what we have found.

 
In Weinberg's scheme\cite{Weinberg:1990rz, Weinberg:1991um}, one
\textit{assumes} the NDA power counting in making the ``effective
potential,'' then treats it \textit{nonperturbatively}. As we have
explained, however, the NDA is the power counting near the trivial fixed
point, where all the operators are irrelevant and should be treated
perturbatively. The so-called infrared enhancement due to two-nucleon
reducible diagrams somehow remedies the above-mentioned mismatch in the
treatment; it makes the ``effective potential'' more relevant. The way
is not systematic however, and in some cases
inconsistent\cite{Kaplan:1996xu}.


The scheme due to Kaplan, Savage, and Wise\cite{Kaplan:1998tg,
Kaplan:1998we} is important because they emphasized one of the most
important aspects of the two-nucleon system; it is close to the
nontrivial fixed point! (The existence of the nontrivial fixed point
itself had been noticed by Weinberg\cite{Weinberg:1991um}.) It is
however very subtle why the KSW scheme succeeds in some channels, and
fails in others. To understand the points, we have to examine it very
carefully.


It is difficult to define the KSW power counting without mentioning the
\textit{Power Divergence Subtraction} (PDS) scheme\footnote{We do not
claim that PDS is indispensable for the KSW power counting, but that the
PDS scheme provides the most unambiguous definition of the KSW power
counting.}. In PDS, the poles at $D=3$ as well as $D=4$ are
subtracted. In the pionless NEFT, there is no pole at $D=4$, so that
\textit{one subtracts as if one were in $D=3$ dimensions}. This is the
heart of the PDS scheme; it changes the (canonical) dimensions of the
operators! For example, the operator $\mathcal{O}^{(S)}_0$ has the
canonical dimension $6$ in $D=4$ dimensions, but would have $4$ in $D=3$
dimensions. In PDS, it is treated as an operator of dimension $4$. This
shift of the dimension by two coincides with the correct anomalous
dimension of the operator, as we found in the previous
section. (Remember that the leading contact operators have anomalous
dimension $2$, and according to Birse \textit{et al.}, all the
``energy-dependent perturbations'' seem to have the same anomalous
dimensions.) We believe that it is the true reason why the KSW/PDS
scheme works near the nontrivial fixed point\footnote{From this point of
view, it is legitimate to think that the KSW/PDS scheme should be
applied only to the case near the nontrivial fixed point, though the RG
equations (\ref{KSWRGE}) know about the trivial fixed point too.}.


Actually, this ``shift by two'' applies to \textit{all} the four-nucleon
operators in the KSW/PDS scheme\footnote{The redundant operators scales
differently than the others, but they are not necessary with dimensional
regularization.}. To see this, consider the KSW/PDS RG equation for the
dimensionless couplings,
\begin{equation}
 \mu\frac{d\gamma_{2n}}{d\mu}=(2n+1)\gamma_{2n}
  +\sum_{m=0}^n\gamma_{2m}\gamma_{2(n-m)},
  \label{KSWRGE}
\end{equation}
where $\gamma_{2n}\equiv (M\mu^{2n+1}/4\pi)C_{2n}$, with $C_{2n}(\mu)$
being given in Eq.~(2.10) in Ref.~\cite{Kaplan:1998we}. Near the nontrivial
fixed point $(\gamma_0^\star, \gamma_2^\star, \gamma_4^\star, \cdots) =
(-1, 0, 0, \cdots)$, the RHS of (\ref{KSWRGE}) becomes
\begin{equation}
 (2n+1)\gamma_{2n}+2\gamma_0\gamma_{2n}
  + \cdots
  \simeq (2n-1)\gamma_{2n},
\end{equation}
where the ellipsis stands for terms quadratic in $\gamma$'s and
independent of $\gamma_0$, which do not contribute to the anomalous
dimensions. This explicitly shows the shift from $(2n+1)$ to
$(2n-1)$\footnote{There is a subtle point about the scaling of the
coupling constants. If one solves the RG equations near the fixed point,
the coupling constant scales $C_{2n} \sim 4\pi / M \Lambda_0^{2n-1}
\mu^{2}$, but it is different from the KSW scaling $C_{2n}\sim 4\pi / M
\Lambda_0^n \mu^{n+1}$. It is due to the assumption that the effective
range is of natural size, $r\sim 1/\Lambda_0$. Namely, the KSW scaling
is not that for a point close to the fixed point, but for a point close
to the critical surface.}.


As far as all the anomalous dimensions are two, the KSW/PDS scheme would
work perfectly. The problem in the pionful NEFT seems that the singular
interactions due to pion exchange in the ${}^3S_1$--${}^3D_1$ channel
would modify the anomalous dimensions of the contact operators strongly,
making the ``shift by two'' rule invalid. Actually,
Birse\cite{Birse:2005um} found that the anomalous dimensions shift
additionally by \textit{minus one half} at the nontrivial fixed point in
the presence of the tensor force due to one-pion exchange. The
investigation in this direction is currently in progress\cite{HKY}.


Several authors considered power countings which are compatible with the
KSW power counting \textit{without} using PDS so that one might think
that the KSW scheme may be defined independent of PDS. In fact, it
appears possible to do so if one \textit{assumes} the scaling of the
coupling constants which is a consequence of the PDS
regularization/renormalization. In Ref.~\cite{vanKolck:1998bw}, van
Kolck introduced a scale $\aleph$, which accounts for the fine-tuning,
and \textit{assumed} the scaling of the coupling constants without
specifying the regularization/renormalization. The scale $\aleph$ seems
to correspond to our $c\Lambda_0$ with $c \ll 1$ introduced in
Eq.~(\ref{singletvariation}). It is important to note that $\aleph$ is a
\textit{physical} scale so that there is no notion of RG with respect to
$\aleph$ in his approach. If one nevertheless identifies $\aleph$ with
the scale $\mu$ in the KSW scheme, the assumed scaling of the couplings
is just the one in the KSW/PDS scheme. On the other hand, using momentum
space subtraction renormalization, Mehen and Stewart\cite{Mehen:1998zz,
Mehen:1998tp} \textit{imposed} the renormalization condition (Eq.~(6) in
Ref.~\cite{Mehen:1998zz})
\begin{equation}
 \left.iA^{(m-1)}\right|_{p=i\mu_R,m_\pi=0}=-iC_{2m}(\mu_R)
  \left(i\mu_R\right)^{2m},\label{MS_renorm}
\end{equation}
on the (part of the) amplitude that scales as $Q^{m-1}$, where $Q$ is a
typical momentum scale. (This power counting for the amplitudes is also
assumed in accordance with the KSW/PDS scheme.) Putting $ \mu_R \sim Q$,
it requires the coupling constant $C_{2m}(\mu_R)$ to scale as $\sim
Q^{m-1} \left(\mu_R\right)^{-2m} \sim Q^{-(m+1)}$, which is essentially
equivalent to the KSW/PDS power counting. It is still unclear to us why
the renormalization condition (\ref{MS_renorm}) reflects the situation
that we are close to the nontrivial fixed point.


Let us finally comment on the confusion in the literature over the scale
$\mu$ introduced in dimensional regularization.  It is important to note
that the scale $\mu$ does not have intrinsic meaning of ``typical
scale.'' It is completely arbitrary and physics is independent of it. In
the usual applications of dimensional regularization to relativistic
field theory, $\mu$ appears in logarithms as $\ln \left( Q^2/\mu^2
\right)$, where $Q$ is the typical scale of the process. It is this
dependence that makes it useful to think $\mu\sim Q$ in order to
suppress the ``large logarithms.'' In our application to NEFT, however,
there is no logarithm, thus no reason to think $\mu\sim Q$.

\section{Conclusion}
\label{sec:summary}


In this paper we studied the Wilsonian RG flow of pionless NEFT using
the Legendre flow equation in order to determine the correct power
counting. We emphasized that the determination of power counting should
be based on the scaling dimensions, because power counting is in essence
the order of magnitude estimate based on dimensional analysis, and the
scaling dimensions are quantum dimensions of the operators.


The RG flow has a nontrivial fixed point as well as a trivial one in
each of the ${}^1S_0$ and ${}^3S_1$--${}^3D_1$ channels. Near the
nontrivial fixed point, we found a relevant operator, the coupling
constant of which is proportional to the inverse of the scattering
length. The nontrivial fixed point is on the boundary of two phases, the
weak coupling and the strong coupling phases. The inverse of the
scattering length is regarded as the order parameter, which takes zero
at the boundary. The real two-nucleon systems are considered to be close
to the nontrivial fixed point.


The difference between the ${}^1S_0$ and ${}^3S_1$--${}^3D_1$ channels
comes from the difference of the phases they are in.  In the strong
coupling phase, the scattering length is positive and the four-nucleon
amplitude has a physical pole which represent a bound state, the
deuteron in the ${}^3S_1$--${}^3D_1$ channel, while in the weak coupling
phase, the scattering length is negative and there is no bound state;
this phase describes the ${}^1S_0$ channel. We believe that this way of
viewing the difference between the two channels is new, and we expect
that it would give some further insight into the character of nuclear
force.


The relation to the work by Birse \textit{et al.} was clarified. The use
of fully developed EFT framework was emphasized.  We also made comments
on other power counting schemes and tried to understand why they succeed
in some cases and fail in others from our point of view. In particular,
we claimed that the true reason why the KSW/PDS scheme works is
identified as the ``shift by two,'' namely, PDS happens to capture the
essential feature that the anomalous dimensions of the contact
two-nucleon interactions are two.


It was essential to include the so-called redundant operators to
consistently perform the Wilsonian RG analysis\cite{Harada:2005tw} in a
fully field theoretical manner. They are also necessary to renormalize
the off-shell amplitudes. The inclusion of them seems essential to
consider the tree-nucleon systems. 


The RG equations derived from the Legendre flow equation are much
simpler than those from the amplitude, while the important physical
information (the existence and the location of fixed points, the global
structure of the flows, the scaling dimensions, etc.) is the same. The
use of the Legendre flow equation would help us to tackle more
difficult problems, where the direct method seems infeasible.


It is very important to apply the present method based on the Wilsonian
RG to determine the power counting of the pionful NEFT, where the power
counting issues are most controversial. The singular tensor force would
change the anomalous dimension of the lowest-order contact operator
drastically in the ${}^3S_1$--${}^3D_1$ channel, invalidating the
``shift by two'' rule. One might imagine that, at a high energy scale,
there is actually no difference of the values of couplings between the
${}^1S_0$ and ${}^3S_1$--${}^3D_1$ channels, but the pion interactions
make them so different. This scenario appears attractive from the
approximate spin-flavor symmetry point of view\cite{Kaplan:1995yg,
Kaplan:1996rk, Mehen:1999qs}. Note that the nontrivial fixed point is
spin-flavor symmetric.


One of the interesting results in this paper is the explicit relation
between the Wilsonian RG analysis and the existence of the bound state
in the strong coupling phase. It would open a new possibility of
studying bound states using the Wilsonian RG. On the other hand, it
would become important when we study the tree-nucleon systems
\textit{without introducing a ``dibaryon'' field }\cite{Kaplan:1996nv,
Bedaque:1997qi, Bedaque:1998mb}. Note that most of the existing
literature on the NEFT study of three-nucleon systems is based on the
``dibaryon'' formalism.  Because our formulation is fully field
theoretical, it is free from the off-shell ambiguities which the
approaches based on potentials suffer from.

\begin{acknowledgments}
 We would like to thank Y.~Yamamoto for several useful discussions.
 We are also grateful to H.~Gies for pointing out the wrong placing of
 the derivative $\tilde{\der}_t$ in Sec.~\ref{sec:Legendre} in the
 earlier manuscript.
 One of the authors (K.H.) is partially supported by Grant-in-Aid for
 Scientific Research on Priority Area, Number of Area 763, ``Dynamics of
 Strings and Fields,'' from the Ministry of Education, Culture, Sports,
 Science and Technology, Japan. 
\end{acknowledgments}

\appendix

\section{IR cutoff function with finite $n$}
\label{sec:ir}

In this Appendix, we give a useful formula involving the IR cutoff
function, and the dependence of the fixed points and the scaling
dimensions of the RG equations on $n$.

\subsection{Evaluation of a typical loop integral}
\label{sec:formulae}

A typical loop integral with the IR cutoff function
$R_{\Lambda}^{(1)}(\bfk^2)$ which appears repeatedly in the derivation
of our RG equations is
\begin{equation}
 I_{m,n}\equiv\int \frac{d^3k}{(2\pi)^3}\tilde{\partial}_t
  \frac{\left(\bfk^{2}\right)^m}
  {A(P) - \frac{\bfk^2}{M} +2R_{\Lambda}^{(1)}(\bfk^2)},
\end{equation}
where  $A(P)$ is a Galilean invariant quantity,
\begin{equation}
 A(P)\equiv P^0-\frac{\bfP^2}{4M},
\end{equation}
and $P^\mu=\left(P^0,\bfP\right)$ is the momentum flowing in (and out
of) the diagram, $P=p_1+p_2 = p_3+p_4$, where $p_1$ and $p_2$ are
initial momenta, and $p_3$ and $p_4$ are final momenta of the
nucleons. The IR cutoff function is chosen as (\ref{cutoffR}) for
definiteness.

Switching to the polar coordinate, making the change of variable to $x
= k/\Lambda$, and expanding the integrand with respect to
$\tilde{A}(P)\equiv \left(M/\Lambda^2\right)A(P)$, we have
\begin{eqnarray}
 I_{m,n}&=&-\frac{n}{\pi^2}M\Lambda^{2m+1}
  \int_0^{\infty}dx x^{2n+2m+4}\exp\left(x^{2n}\right)
  \left[\frac{\hat{R}(x)}{\tilde{A}(P)-x^2
   \left(1-\hat{R}(x)\right)}\right]^2
  \nonumber\\
 &=&-\frac{M\Lambda^{2m+1}}{2\pi^2}
  \left[
   \Gamma\left(1+\frac{2m+1}{2n}\right)
   +\tilde{A}(P)\Gamma\left(1+\frac{2m-1}{2n}\right)
   \left(2-2^{\frac{1-2m}{2n}}\right)+\mathcal{O}(\tilde{A}^2)
  \right],
  \nonumber \\
\end{eqnarray}
where $\hat{R}(x)\equiv \left(1-\exp \left(x^{2n}\right)\right)^{-1}$
has been introduced. In the present paper, we disregard the terms of
$\mathcal{O}(\tilde{A}^2)$ as higher orders.

We may decompose $A(P)$ into the following Galilean invariant quantities,
\begin{eqnarray}
 A(P)&=&\left(p_1^0-\frac{\bfp_1^2}{2M}\right)
  +\left(p_2^0-\frac{\bfp_2^2}{2M}\right)
  +\frac{1}{4M}\left(\bfp_1-\bfp_2\right)^2.
\end{eqnarray}
Combining possible multiplicative momentum factors outside the integral,
we can identify the contributions to the various operators in our theory
space.  For example, contributions from the first two terms renormalize
a redundant operator and the last term does the operators which contain
spatial derivatives.

\subsection{Dependence on the IR cutoff function}
\label{sec:n-dep}

The RG equations are derived and solved with a finite value of $n$. In
this section, we show the $n$ dependence of the fixed points and the
scaling dimensions.

The dependence on $n$ might give some useful information on the behavior
of the present approximation. Note that if no approximation were made,
the universal quantities such as the scaling dimension would not depend
on how the RG transformation is defined. Weak dependence would suggest
that the approximation is good. We will see shortly that the most of the
fixed points as well as the scaling dimensions have very weak
dependence. The fixed point with strong dependence on $n$ may be
regarded as ``spurious'' and, for other fixed points, the approximation
is very good.

In general, a sharp momentum cutoff gives a poor convergence and a
smooth cutoff function is preferred\cite{Litim:2001dt}. In the present
case, however, the dependence on $n$ is very weak and the use of a sharp
cutoff is expected to be reasonable. In addition, it gives the simplest
expressions.

For a finite value of $n$, we find four fixed points including the
trivial one. The dependence of the three nontrivial fixed points in the
${}^1S_0$ channel is given in Tables~\ref{tab:table1}, \ref{tab:table2},
and \ref{tab:table3}. (The dependence is similar for the
${}^3S_1$--${}^3D_1$ channel.) The last column shows the three
eigenvalues of the linearized RG equations at the fixed point.  Let us
call these fixed points $A$, $B$, and $C$.

Note that, although we have an analytic expression for the RG equations
for any $n$, the actual evaluation is done with the aid of a
computer. We use \textsl{Mathematica}, but some numerical errors are
not avoidable.

One can easily see that the fixed points $A$ and $B$ have little $n$
dependence (for $n\ge 10$), while the fixed point $C$ depends on $n$
strongly. We disregard the fixed point $C$. Note that even for the
smallest integer value of $n$, $n=2$ ($n=1$ does not satisfy the
condition (\ref{ir-func}).), the scaling dimensions for the fixed point
$A$ are the same while the fixed points $B$ and $C$ disappear there.

The fixed point $B$ has complex eigenvalues, though they have weak
dependence on $n$. As discussed in the main text, the eigen-operators do
not have the definite scaling properties there and we disregard it too
in our discussion of the power counting. At the moment we know neither
if it is just an artifact, nor, even if it is not, its physical
implications.

\begin{table*}
\caption{\label{tab:table1}The dependence on $n$ of the fixed point $A$.}
\begin{ruledtabular}
\begin{tabular}{ccc}
$n$&location of the fixed point($x^{\star},y^{\star},z^{\star}$)&scaling dimensions\\
\hline 
$2$&$(-1.10326, -0.60467, 0.60467)$&$(-2,-1,1)$\\
$10$&$(-1.02722,-0.524991,0.524991)$&$(-2,-1,1)$\\
$10^2$&$(-1.00287,-0.502587,0.502587)$&$(-2,-1,1)$\\
$10^3$&$(-1.00029,-0.50026,0.50026)$&$(-2,-1,1)$\\
$10^4$&$(-1.00003,-0.599926,0.500026)$&$(-2,-1,1)$\\
$\infty $&$(-1,-\frac{1}{2},\frac{1}{2})$&$(-2,-1,1)$\\
\end{tabular}
\end{ruledtabular}
\end{table*}

\begin{table*}
\caption{\label{tab:table2}The dependence on $n$ of the fixed point $B$.}
\begin{ruledtabular}
\begin{tabular}{ccc}
$n$&location of the fixed point($x^{\star},y^{\star},z^{\star}$)&scaling dimensions\\
\hline
$2$& Does not exist & Does not exist \\
$10$&$(-9.85064,8.33328,-1.27074)$&$(-0.79483\pm 2.18601 i,3)$\\
$10^2$&$(-9.03165,7.52438,-1.46746)$&$(-0.502903\pm 2.3961 i,3)$\\
$10^3$&$(-9.00266,7.50188,-1.49663)$&$(-0.50003\pm 2.3979 i,3)$\\
$10^4$&$(-9.00026,7.50018,-1.49966)$&$(-0.494178\pm 2.4005 i,3.00465)$\\
$\infty $&$(-9,\frac{15}{2},-\frac{3}{2})$&
$(-\frac{1}{2}\pm i \frac{\sqrt{23}}{2},3)$\\
\end{tabular}
\end{ruledtabular}
\end{table*}

\begin{table*}
\caption{\label{tab:table3}The dependence on $n$ of the fixed point $C$.}
\begin{ruledtabular}
\begin{tabular}{ccc}
$n$&location of the fixed point($x^{\star},y^{\star},z^{\star}$)&scaling dimensions\\
\hline
$2$& Does not exist & Does not exist \\
$10$&$(-71.9608,77.804,-5.93769)$&$(-62.929,0.874861,3)$\\
$10^2$&$(-6226.88,6961.68,-702.804)$&$(-6202.06,0.99871,3)$\\
$10^3$&$(-609432,681363,-71583.4)$&$(-609248,0.999984,3.81652)$\\
$10^4$&$(-6.08078\times 10^7,6.79826\times 10^7,-7.17129\times 10^6)$&
$(-6.0806\times 10^7,1,3)$\\
$\infty $&Does not exist&Does not exist\\
\end{tabular}
\end{ruledtabular}
\end{table*}

\section{Explicit derivation of RG equations from the amplitudes}
\label{sec:amplitudes}

In this Appendix, we consider the (off-shell) two-nucleon amplitudes in
the ${}^3S_1$--${}^3D_1$ channel in order to derive the RG equations.
(The amplitude in the ${}^1S_0$ channel is given in
Ref.~\cite{Harada:2005tw}.)  The amplitudes take a matrix form due to
the presence of the ${}^3S_1$--${}^3D_1$ mixing and satisfy the
following Lippmann-Schwinger equation in the center-of-mass frame,
\begin{eqnarray}
 -i\mathcal{A}_{ij}\left(p_0,\bfp_2^2, \bfp_1^2\right)
  &=&-iV_{ij}\left(p_0,\bfp_2^2, \bfp_1^2\right)
  \nonumber \\
 &&{}+\int \frac{d^3 k}{(2\pi)^3}
  \left(-iV_{ik}\left(p_0,\bfp_2^2, \bfk^2\right)\right)
  \frac{i}{p^0-\frac{\bfk^2}{2M}+i\epsilon}
  \left(-i\mathcal{A}_{kj}\left(p_0,\bfk^2, \bfp_1^2\right)\right),
  \nonumber \\
\end{eqnarray}
where the subscript labels the partial waves, $i,j,k=\{S,D\}$, and
$V_{ij}$ stands for the vertex in momentum space.  The momentum integral
is taken over $0\le k <\Lambda$.

We take the averaged action (\ref{truncation}) as our action, and write
the vertex $V_{ij}$ to order $\mathcal{O}(p^2)$,
\begin{eqnarray}
 V_{SS}&=&C_0^{(T)}+4C_2^{(T)}\left(\bfp_1^2+\bfp_2^2\right)
  -2B^{(T)}\left(p^0-\frac{\bfp_1^2+\bfp_2^2}{2M}\right),\\
 V_{SD}&=&\frac{4\sqrt{2}}{3}C_2^{(SD)}\bfp_1^2,\quad
  V_{DS}=\frac{4\sqrt{2}}{3}C_2^{(SD)}\bfp_2^2,\quad
  V_{DD}=0,
\end{eqnarray}
where $p^0$ is the center-of-mass energy, $\bfp_1$ and $\bfp_2$ are the
relative momenta in the initial and final states respectively. We
here suppress the spin and isospin dependence. See the appendix in
Ref.~\cite{Fleming:1999ee} for a useful technique.

We consider the following ansatz,
\begin{subequations}
\begin{eqnarray}
 \mathcal{A}_{SS}(p^0, \bfp_2^2, \bfp_1^2)
  &=&
  T_1(p_0)+T_2(p_0)\left(\bfp_1^2+\bfp_2^2\right)
  +T_3(p_0)\bfp_1^2 \bfp_2^2, \\
 \mathcal{A}_{SD}(p^0, \bfp_2^2, \bfp_1^2)
  &=&
  T_4(p_0)\bfp_1^2+T_5(p_0)\bfp_1^2 \bfp_2^2,\\
 \mathcal{A}_{DS}(p_0, \bfp_2^2, \bfp_1^2)
  &=&
  T_6(p_0)\bfp_2^2 + T_7(p_0) \bfp_1^2 \bfp_2^2,\\
 \mathcal{A}_{DD}(p_0, \bfp_2^2, \bfp_1^2)
  &=&
  T_8(p_0) \bfp_1^2 \bfp_2^2,
\end{eqnarray}
\end{subequations}
and obtain the solution,
\begin{subequations}
\begin{eqnarray}
 T_1&=&\frac{1}{D}
  \left[
   \left(
    C_0^{(T)}-2B^{(T)}p^0
   \right)
   + \left\{
      \left(
       4C_2^{(T)}+\frac{B^{(T)}}{M}
      \right)^2
      +\frac{32}{9}\left(C_2^{(SD)}\right)^2
     \right\}I_2
  \right],\\
 T_2&=&\frac{1}{D}
 \left(
  4C_2^{(T)}+\frac{B^{(T)}}{M} 
 \right)
 \left[
  1-\left(
     4C_2^{(T)}+\frac{B^{(T)}}{M} 
    \right)I_1
 \right],\\
 T_3&=&\frac{1}{D}
 \left(
  4C_2^{(T)}+\frac{B^{(T)}}{M}
 \right)^2I_0, \\
 T_4&=&T_6=\frac{1}{D}
 \frac{4\sqrt{2}}{3}C_2^{(SD)}
 \left(
  1-\left(4C_2^{(T)}+\frac{B^{(T)}}{M}\right)I_1
 \right),\\
 T_5&=&T_7=\frac{1}{D}
 \frac{4\sqrt{2}}{3}C_2^{(SD)}
 \left(
  4C_2^{(T)}+\frac{B^{(T)}}{M}
 \right)I_0, \\
 T_8&=&\frac{32}{9} \frac{1}{D}
 \left(
  C_2^{(SD)}
 \right)^2I_0,
\end{eqnarray}
\end{subequations}
with
\begin{eqnarray}
 D&=&1-\left(C_0^{(T)}-2B^{(T)}p^0\right)I_0
  -2\left(4C_2^{(T)}+\frac{B^{(T)}}{M}\right)I_1
  +\left(4C_2^{(T)}+\frac{B^{(T)}}{M}\right)^2I_1^2
  \nonumber\\
 &&-\left(
     \left(4C_2^{(T)}+\frac{B^{(T)}}{M}\right)^2
     +\frac{32}{9}\left(C_2^{(SD)}\right)^2
    \right)I_0I_2,
\end{eqnarray}
where $I_n$ are integrals defined by 
\begin{equation}
 I_n=-\frac{M}{2\pi^2}
  \int^{\Lambda}dk\frac{k^{2n+2}}{k^2+\mu^2},\quad
  \mu=\sqrt{-Mp^0-i\epsilon}.
\end{equation}
The RG equations for the couplings are obtained by requiring that the
inverse of the off-shell amplitudes ($\mathcal{A}_{SS},
\mathcal{A}_{SD}$ in this case) are independent of the cutoff $\Lambda$,
in the expansion in powers of $\bfp_1^2$, $\bfp_2^2$, and $\mu^2$.  We
introduce dimensionless coupling constants as (\ref{dimensionlesstriplet}).

The resulting RG equations are written as
\begin{subequations}
 \label{ampRGEtriplet}
\begin{eqnarray}
\Lambda\frac{dX}{d\Lambda}
 &=&
 \frac{1}{X}\left(X-1\right) \left( Y+3X^2 \right),
 \\
 \Lambda\frac{dY}{d\Lambda}
  &=&
  \frac{Y}{X^2}\left( 6X^3-5X^2+2XY-Y\right),
  \\
 \Lambda\frac{dZ}{d\Lambda}
  &=&
  \frac{1}{X^2}\left( 6X^3Z-3X^2Z+2XYZ+Y^2\right),
  \\
 \Lambda\frac{dw'}{d\Lambda}
  &=&
  \frac{1}{X}\left( 3X^2w'+Yw' \right),
\end{eqnarray}
\end{subequations}
where we have introduced $X = 1 + \left( y'+z' \right)/3$,
$Y = x' - \left(\left(y'+z'\right)^2+2w'^2\right)/5$, and
$Z = 2y'+\left(\left(y'+z'\right)^2-2w'^2\right)/3$. Compare them with
those obtained in Ref.~\cite{Harada:2005tw} for the ${}^1S_0$
channel. They are identical if one sets $w'=0$.

Although they look complicated and very different from
Eqs.~(\ref{RGEtriplet}), when expanded in couplings, they agree with
Eqs.~(\ref{RGEtriplet}) to the second order.

These RG equations have only two fixed points (as far as
\textsl{Mathematica} can find); a trivial fixed point
$\left( X^{\star}, Y^{\star}, Z^{\star}, w^{\star}\right) = \left( 1, 0,
0, 0 \right)$ and a nontrivial fixed point $\left(X^{\star}, Y^{\star},
Z^{\star}, w^{\star} \right) = \left(1, -1, -1, 0 \right)$ which
corresponds to $\left( x'^{\star}, y'^{\star}, z'^{\star}, w'^{\star}
\right) = \left(-1, -1/2, 1/2, 0\right)$. The locations of these fixed
points are the same as those obtained from Eqs.~(\ref{RGEtriplet}).

Substituting
\begin{equation}
x'=x'^{\star}+\delta x',\quad
y'=y'^{\star}+\delta y',\quad
z'=z'^{\star}+\delta z',\quad
w'=w'^{\star}+\delta w',
\end{equation}
into (\ref{ampRGEtriplet}), we obtain the linearized RG equations at the
nontrivial fixed point,
\begin{eqnarray}
 \frac{d}{dt}\left(
	      \begin{array}{c}
	       \delta x'\\
	       \delta y'\\
	       \delta z'\\
	       \delta w'
	      \end{array}
	     \right)=\left(
	     \begin{array}{cccc}
	       1& 2& 2&0\\
	       2& \frac{2}{3}& \frac{5}{3}&0\\
	      -2&-\frac{8}{3}&-\frac{11}{3}&0\\
	      0&0&0&-2
	     \end{array}
		     \right)
	     \left(
	      \begin{array}{c}
	       \delta x'\\
	       \delta y'\\
	       \delta z'\\
	       \delta w'
	      \end{array}
       \right),
\end{eqnarray}
where we have introduced $t=\ln \left(\Lambda_0/\Lambda\right)$.  

These equations can be solved easily,
\begin{eqnarray}
\left(
 \begin{array}{c}
  \delta x^{\prime}\\
  \delta y^{\prime}\\
  \delta z^{\prime}\\
  \delta w^{\prime}
 \end{array}
\right)
&=&
a_1^{\prime}
\left(
 \begin{array}{c}
  2\\
  1\\
  -4\\
  0
 \end{array}
\right)\left(\frac{\Lambda}{\Lambda_0}\right)^2
+a_2^{\prime}
\left(
 \begin{array}{c}
  0\\
  0\\
  0\\
  1
 \end{array}
\right)\left(\frac{\Lambda}{\Lambda_0}\right)^2
\nonumber\\
 &&{}
+b^{\prime}
\left(
 \begin{array}{c}
  0\\
  -1\\
  1\\
  0
 \end{array}
\right)\left(\frac{\Lambda}{\Lambda_0}\right)
+c^{\prime}
\left(
 \begin{array}{c}
  1\\
  1\\
  -1\\
  0
 \end{array}
\right)\left(\frac{\Lambda_0}{\Lambda}\right),
\end{eqnarray}
where $a_1^{\prime}$, $a_2^{\prime}$, $b^{\prime}$, and $c^{\prime}$ are
dimensionless constants. This corresponds to Eq.~(\ref{singletvariation}).

We obtain the renormalized off-shell amplitudes near the nontrivial
fixed point to the second order in powers of $\bfp^2$ and $p^0$ by
substituting the solutions of linearized RG equations, for example,
\begin{equation}
\left.\mathcal{A}^{-1}_{SS}(p_0,{\bfp^{\prime}}^2,\bfp^2)\right|_*
 = -\frac{M}{4\pi}
 \left[
  \frac{2}{\pi}c'\Lambda_0
  -\frac{4}{\pi}b'\left(\frac{Mp_0}{\Lambda_0}\right)
  -i\sqrt{Mp_0}
 \right]+\cdots,
\end{equation}
where the ellipsis denotes terms of higher orders of $\bfp^2/\Lambda_0$
and $p^0/\Lambda_0$.  We notice that $a_1^{\prime}$ and $a_2^{\prime}$
do not contribute to the inverse of the amplitude to this order.  The
existence of such eigenvectors whose coefficients do not show up in the
on-shell amplitudes is related to the equivalence theorem considered in
the previous paper\cite{Harada:2005tw}. Actually, the equivalence
relation (3.21) of Ref.~\cite{Harada:2005tw} gives ``trajectories'' in
the space of couplings, and the tangential direction at the nontrivial
fixed point is the direction of the eigenvector whose coefficient does
not appear in the physical amplitude. This would suggest that all the
``momentum-dependent perturbations'' discussed in
Ref.~\cite{Birse:1998dk} form ``surfaces'' in the space of couplings
near the fixed point on which any two points are related by some
equivalence relation.


\bibliography{NEFT,NPRG}

\end{document}